\documentclass[ALICE,manyauthors]{cernphprep}
\usepackage[english]{babel}
\usepackage{graphicx}
\usepackage{verbatim}
\usepackage{amsfonts}
\usepackage{makeidx}
\usepackage{color}
\usepackage{amsmath}
\usepackage{lineno}
\usepackage{epsfig}
\usepackage{epstopdf}
\usepackage{hyperref}
\usepackage{cite}
\newcommand{\sqrtsNN}{\sqrt{s_{\rm \scriptscriptstyle NN}}}

\newcommand{\gsim}{\,{\buildrel > \over {_\sim}}\,}
\newcommand{\av}[1]{\left\langle #1 \right\rangle}

\newcommand{\MeV}{\mathrm{MeV}}
\newcommand{\GeV}{\mathrm{GeV}}
\newcommand{\TeV}{\mathrm{TeV}}

\newcommand{\mev}{\mathrm{MeV}}
\newcommand{\gev}{\mathrm{GeV}}

\newcommand{\tev}{\mathrm{TeV}}

\newcommand{\mum}{\mathrm{\mu m}}

\newcommand{\PbPb}{\mbox{Pb--Pb}}

\newcommand{\RAA}{R_{\rm AA}}
\newcommand{\TAA}{T_{\rm AA}}
\newcommand{\pt}{p_{\rm T}}
\newcommand{\mt}{m_{\rm T}}

\newcommand{\Dstophipi}{{\rm D_s^{+}\to \phi\pi^+}}

\newcommand{\DstophipitoKKpi}{{\rm D_s^{+}\to \phi\pi^+\to K^-K^+\pi^+}}
\newcommand{\DplustoKKpi}{{\rm D^{+}\to K^-K^+\pi^+}}
\newcommand{\phitoKK}{{\rm \phi\to  K^-K^+}}
\newcommand{\DstoKzerostarK}{{\rm D_s^{+}\to \overline{K}^{*0} K^+}}
\newcommand{\Dstofzeropi}{{\rm D_s^{+}\to f_{0}(980) \pi^+}}
\newcommand{\fzero}{{\rm f_{0}(980)}}
\newcommand{\Kzerostar}{{\rm \overline{K}^{*0}}}
\newcommand{\Dzero}{{\rm D^0}}

\newcommand{\Dstar}{{\rm D^{*+}}}

\newcommand{\Dplus}{{\rm D^+}}

\newcommand{\Ds}{{\rm D_s^+}}
\newcommand{\Dsminus}{{\rm D_s^-}}
\newcommand{\Dspm}{{\rm D_s^\pm}}
\newcommand{\KKpi}{{\rm K^-K^+\pi^+}}

\newcommand{\Lxy}{L_{xy}}
\newcommand{\dEdx}{{\rm d}E/{\rm d}x}

\newcommand{\costhetap}{\cos \theta_{\rm point}}
\newcommand{\costhetapxy}{\cos \theta^{xy}_{\rm point}}

\begin{document}

\begin{titlepage}
\PHyear{2015}
\PHnumber{253}      
\PHdate{14 September}  
%

\title{Measurement of D$\mathbf{\mathrm{_s^+}}$ production and nuclear modification factor \\ in Pb--Pb collisions at $\mathbf{\sqrtsNN=2.76}$~TeV}
\ShortTitle{D$_{\rm s}$ production in Pb--Pb collisions}

\Collaboration{ALICE Collaboration\thanks{See Appendix~\ref{app:collab} for the list of collaboration members}}
\ShortAuthor{ALICE Collaboration} 

\begin{abstract} 
The production of prompt $\Ds$ mesons was measured 
for the first time in collisions of heavy nuclei with the ALICE
detector at the LHC.
The analysis was performed on a data sample of Pb--Pb collisions at a
centre-of-mass energy per nucleon pair, $\sqrtsNN$, of $2.76~\tev$ in two 
different centrality classes, namely 0--10\% and 20--50\%.
$\Ds$ mesons and their antiparticles were reconstructed at mid-rapidity 
from their hadronic decay channel $\Dstophipi$, 
with $\phitoKK$, 
in the transverse momentum intervals $4<\pt<12~\GeV/c$ and 
$6<\pt<12~\GeV/c$ for the 0--10\% and 20--50\% centrality classes, respectively. 
The nuclear modification factor $\RAA$ was computed by comparing the
 $\pt$-differential production yields in Pb--Pb collisions to those in 
proton--proton (pp) collisions at the same energy. This pp reference was obtained 
using the cross section measured 
at $\sqrt{s}= 7~\tev$ and scaled to $\sqrt{s}= 2.76~\tev$.
The $\RAA$ of $\Ds$ mesons was compared to that of non-strange D mesons
in the 10\% most central Pb--Pb collisions.
At high $\pt$ ($8<\pt<12~\gev/c\,$) a suppression of the $\Ds$-meson yield
by a factor of about three, compatible within uncertainties with that of 
non-strange D mesons, is observed. 
At lower $\pt$ ($4<\pt<8~\gev/c\,$) the values of the $\Ds$-meson 
$\RAA$ are larger than those of non-strange D mesons, although compatible
within uncertainties.
The production ratios $\Ds/\Dzero$ and $\Ds/\Dplus$ were also measured in
Pb--Pb collisions and compared to their values in proton--proton collisions.
\end{abstract}

\end{titlepage}

\setcounter{page}{2}

\section{Introduction}
\label{sec:intro}
Calculations of Quantum Chromodynamics (QCD) on the lattice
predict that strongly-interacting matter at temperatures
exceeding the pseudo-critical value of about $T_{\rm c} \approx 145-165~\mev$
and vanishing baryon density behaves as a deconfined Plasma of Quarks and 
Gluons (QGP)~\cite{Borsanyi:2010bp,Bazavov:2014pvz}.
In this state, partons are the relevant degrees of freedom and
chiral symmetry is predicted to be restored.
The conditions to create a QGP are expected to be attained in 
collisions of heavy nuclei at high energies. 
This deconfined state of matter exists for a short time 
(few fm/$c$), during which the medium created in the collision expands and 
cools down until its temperature drops below the pseudo-critical value 
$T_{\rm c}$ and the process of hadronisation takes place.

Heavy quarks (charm and beauty) are sensitive probes to investigate the properties of 
the medium formed in heavy-ion collisions.
They are produced in quark-antiquark pairs predominantly at the initial stage of the collision
in hard-scattering processes 
characterized by timescales shorter than the QGP formation 
time~\cite{Biro93,Vogt2007,Rapp2009}.
The heavy quarks propagate through the expanding hot and dense medium,
thus experiencing the effects of the medium over its entire evolution.
While traversing the medium, they interact with its constituents 
via both inelastic and elastic QCD processes, exchanging energy and momentum
with the expanding medium~\cite{Rapp2009,ADSW}.
For heavy quarks at intermediate and high momentum, these interactions lead to
energy loss due to medium-induced gluon radiation and collisional 
processes.

Evidence for heavy-quark in-medium energy loss is provided by the observation 
of a substantial modification of the transverse momentum ($\pt$) 
distributions of heavy-flavour decay 
leptons~\cite{starHFEraa,PhenixHFE2006,PhenixHFEraav2,ALICEHFMRAA}, 
D mesons \cite{ALICEDRAA,STARD0} and
non-prompt J/$\psi$~\cite{CMSJpsi} in Au--Au and Pb--Pb 
collisions at RHIC and LHC energies as compared to proton--proton (pp) 
collisions.
This modification is usually quantified by the nuclear modification factor
$\RAA$, defined as the ratio between the yield measured in nucleus--nucleus 
collisions and the cross section in pp interactions scaled by the average 
nuclear overlap function.
In absence of nuclear effects, $\RAA$ is expected to be unity.
Parton in-medium energy loss causes a suppression of hadron yields, $\RAA<1$, 
at intermediate and high transverse momentum ($\pt>3~\gev/c$).
In central nucleus--nucleus collisions at RHIC and LHC energies,
$\RAA$ values significantly lower than unity were observed for heavy-flavour 
hadrons with $\pt$ values larger than $3-4~\GeV/c$.
In this $\pt$ range, the D-meson yields measured in p--Pb collisions at 
$\sqrtsNN=5.02$~TeV are consistent with binary-scaled pp cross 
sections~\cite{ALICEDRpPb}, providing clear evidence that the suppression 
observed in Pb--Pb collisions is not due to cold nuclear matter effects
and is induced by a strong coupling of the charm quarks with the hot and 
dense medium.

In case of substantial interactions with the medium, heavy quarks lose
a significant amount of energy while traversing the fireball and may 
participate in the collective expansion of the system and possibly 
reach thermal equilibrium with the medium constituents.
In this respect, the measurement of a positive elliptic flow $v_2$
of D mesons at LHC energies~\cite{ALICEDv2lett,ALICEDv2artic}
and of heavy-flavour decay electrons at RHIC 
energies~\cite{PhenixHFE2006,PhenixHFEraav2,STARv2}
provides an indication that the interactions 
with the medium constituents transfer to charm quarks information on the 
azimuthal anisotropy of the system.

It is also predicted that a significant fraction of low- and 
intermediate-momentum heavy quarks could hadronise via recombination with 
other quarks from the medium~\cite{Molnar,GrecoKoLevai,GrecoKoRapp}.
An important role of hadronisation via (re)combination, either during
the deconfined phase~\cite{Thews2000} or at the phase boundary~\cite{PBM2000}, 
is indeed supported by the results of J/$\psi$ nuclear modification factor and 
elliptic flow at low $\pt$~\cite{ALICEjpsiraa1,ALICEjpsiraa2,ALICEjpsiv2}.
Hadronisation via recombination allows in some models,
e.g.~\cite{Gossiaux2009,TAMU,CaoQinBass}, a better 
description of heavy-flavour production measurements at
RHIC and LHC energies, 
in particular the $\RAA$ of $\Dzero$ mesons 
at low $\pt$ measured in Au--Au collisions at 
$\sqrtsNN = 200~\gev$~\cite{STARD0} and the 
positive and sizable D-meson $v_2$ in 
Pb--Pb collisions at $\sqrtsNN = 2.76~\tev$~\cite{ALICEDv2lett}.

The measurement of $\Ds$-meson production in Pb--Pb collisions can provide 
crucial additional information for understanding the 
interactions of charm quarks with the strongly-interacting 
medium formed in heavy-ion collisions at high energies.
In particular, the $\Ds$-meson yield is sensitive to strangeness production 
and to the hadronisation mechanism of charm quarks.

An enhancement of strange particle production in heavy-ion collisions as
compared to pp interactions was long suggested as a possible signal of QGP 
formation~\cite{RafelskiMuller,KochRafelski}.
Strange quarks are expected to be abundant in a deconfined medium due to 
the short time needed to reach equilibrium values among the parton species 
and to the lower energy threshold for $s\overline{s}$ production.
A pattern of strangeness enhancement increasing with the hadron strangeness 
content when going from pp (p--A) to heavy-ion collisions was observed at 
the SPS~\cite{NA57_158,NA57_40,NA49_Kpi,NA49_LambdaXi}, at
RHIC~\cite{STAR_hyperons} and at the LHC~\cite{ALICE_hyperons}.
In the frame of the statistical hadronisation models, strange particle
production in heavy-ion collisions follows the expectation for a 
grand-canonical ensemble.
In contrast, for pp collisions canonical suppression effects are found to 
be important, reducing the phase space available for strange 
particles~\cite{RedlichHamieh,BraunMunzingerRedlichStachel}.
In this context, the increase in strange particle yields in heavy-ion 
collisions compared to pp interactions is viewed as due primarily to the 
lifting of the canonical suppression. 

This strangeness enhancement effect could also affect the production of 
charmed hadrons if the dominant mechanism for D-meson formation at 
low and intermediate momenta is in-medium hadronisation of charm quarks via 
recombination with light quarks.
Under these conditions, the relative yield 
of $\Ds$ mesons with respect to non-strange charmed mesons at low $\pt$ is predicted to be enhanced
in nucleus--nucleus collisions as compared to pp 
interactions~\cite{Andronic2003,RafelskiKuznetsova,HeFriesRapp}.
The comparison of the $\pt$-differential production yields 
of non-strange D mesons and of $\Ds$ mesons in Pb--Pb and pp 
collisions is therefore sensitive to the role of recombination in charm-quark 
hadronisation.

A consequence of the possibly enhanced production of $\Ds$ mesons in heavy-ion 
collisions would be a slight reduction of the fraction of charm quarks 
hadronising into non-strange meson species.
Therefore, the measurement of the $\Ds$-meson production is also relevant
for the interpretation of the comparison of the nuclear modification
factors of non-strange D mesons and light-flavour 
hadrons (pions)~\cite{ALICEDRAA,Adam:2015sza}, which is predicted 
to be sensitive to the quark-mass and colour-charge dependence of parton 
in-medium energy loss~\cite{ADSW,WHDG,Djordjevic}.
Furthermore, due to this possible modification of the relative abundances
of D-meson species, measuring the $\Ds$ yield at low $\pt$ is needed 
also to determine the total charm production cross section in Pb--Pb 
collisions.

The $\pt$-differential inclusive production cross section of 
prompt\footnote{In this paper, 'prompt' indicates D mesons produced at the 
interaction point, either directly in the hadronisation of the charm quark or 
in strong decays of excited charm resonances.
The contribution from weak decays of beauty hadrons, which
gives rise to feed-down D mesons displaced from the interaction vertex,
was subtracted.}
$\Ds$ mesons (average of particles and antiparticles) was measured in pp 
collisions 
at $\sqrt{s}=7~\tev$ with the ALICE detector and it was found to be described 
within uncertainties 
by perturbative QCD (pQCD) calculations~\cite{ALICEDspp7}.
The $\Ds$ nuclear modification factor was measured in p--Pb collisions 
at $\sqrtsNN=5.02~\TeV$ and found to be consistent with unity~\cite{ALICEDRpPb}.
In this paper, we report on the measurement of prompt $\Ds$-meson production
and nuclear modification factor in Pb--Pb collisions at $\sqrtsNN=2.76~\tev$.
$\Ds$ mesons (and their antiparticles) were reconstructed at mid-rapidity, 
$|y|<0.5$, through their hadronic decay channel 
$\Dstophipi$ with a subsequent decay $\phitoKK$.
The production yield was measured in two classes of collision centrality, 
central (0--10\%) and semi-central (20--50\%), and compared to a
binary-scaled pp reference obtained by scaling the cross section measured at
$\sqrt{s}=7~\tev$ to the Pb--Pb centre-of-mass energy via a pQCD-driven 
approach.
The experimental apparatus and the data sample of Pb--Pb collisions used for 
this analysis are briefly presented in Section~\ref{sec:sample}.
In Section~\ref{sec:selection}, the $\Ds$ meson reconstruction strategy, 
the selection criteria and the raw yield extraction from the KK$\pi$
invariant mass distributions are discussed.
The corrections applied to obtain the $\pt$-differential production yields
of $\Ds$ mesons, 
including the subtraction of the non-prompt contribution from
beauty-hadron decays, are described in Section~\ref{sec:corrections}. 
The various sources of systematic uncertainty are discussed in 
detail in Section~\ref{sec:systematics}.  
The results on the $\Ds$-meson production yield and nuclear modification factor
are presented in Section~\ref{sec:results} together with 
the comparison to non-strange D-meson $\RAA$ and to model calculations.
The $\Ds/\Dzero$ and $\Ds/\Dplus$ yield ratios in three $\pt$ intervals 
for the 10\% most central Pb--Pb collisions are compared to those in pp collisions.

\section{Apparatus and data sample}
\label{sec:sample}
The ALICE detector and its performance are described in detail in 
Refs.~\cite{ALICEjinst} and~\cite{ALICEperf}, respectively.
The apparatus consists of a central barrel covering the pseudorapidity region 
$|\eta|<$ 0.9, a forward muon spectrometer ($-4.0 <\eta< -2.5$) and a set of 
detectors for triggering and event centrality determination. 
The detectors of the central barrel are located inside a 0.5~T magnetic field 
parallel to the LHC beam direction, that corresponds to the $z$-axis in 
the ALICE reference frame. 
The information provided by the following detectors was utilised to perform 
the analysis presented in this paper: the Inner Tracking System (ITS), the 
Time Projection Chamber (TPC) and the Time Of Flight (TOF) detector 
were used to reconstruct and identify charged particles at mid-rapidity, 
while the V0 scintillator detector provided the information for triggering, 
centrality determination and event selection. 
The neutron Zero Degree Calorimeters (ZDC) were also used,
together with the V0 detector, for the event selection.

The trajectories of the D-meson decay particles are reconstructed from
their hits in the ITS and TPC detectors.
Particle identification is performed utilising the information from the 
TPC and TOF detectors.
The ITS consists of six cylindrical layers of silicon detectors
covering the pseudorapidity interval $|\eta|<0.9$. 
The two innermost layers,  located at 3.9 and 7.6 cm from the beam line, 
are composed of Silicon Pixel Detectors (SPD). 
The two intermediate layers are equipped with Silicon Drift Detectors (SDD) 
and the two outermost layers, with a maximum radius of 43.0 cm, are composed 
of double-sided Silicon Strip Detectors (SSD). 
The high spatial resolution of the ITS detectors, together with the low 
material budget ($\sim7.7\%$ of a radiation length at $\eta=0$) and the
small distance from the interaction point, provides a resolution on the 
track impact parameter (i.e.\,the distance of closest approach of the track 
to the primary vertex) better than 65 $\mum$ for transverse momenta 
$\pt>1~\gev/c$ in Pb--Pb collisions~\cite{ALICEperf}. 
The TPC, covering the pseudorapidity interval $|\eta|<0.9$, provides track reconstruction with up to 159 points along the trajectory of a charged particle and allows its identification via the measurement of specific energy loss d$E$/d$x$.
Particle identification is complemented with the particle time-of-flight measured with the TOF detector, which is composed of Multi-gap Resistive Plate Chambers and is positioned at 370-399 cm from the beam axis, covering the full azimuth and the pseudorapidity interval $|\eta|<0.9$.  The TPC and TOF information provides
pion/kaon separation at better than $3\,\sigma$ level for tracks with momentum 
up to $2.5~\gev/c$~\cite{ALICEperf}.

The analysis was performed on a sample of $\PbPb$ collisions 
at centre-of-mass energy per nucleon pair, $\sqrtsNN$, of $2.76~\tev$
collected in 2011. 
The events were recorded with an interaction trigger that required coincident 
signals in both scintillator arrays of the V0 detector, covering
the pseudorapidity ranges $-3.7 < \eta < -1.7$ and $2.8 < \eta < 5.1$, respectively.
An online selection based on the V0 signal amplitude was used to record samples 
of central and semi-central collisions through two separate trigger classes.
Events were further selected offline to remove background from 
beam-gas interactions on the basis of the timing information provided by the 
V0 and the neutron ZDC detectors (two hadronic calorimeters located at 
$z=114$~m on both sides of the interaction point covering the interval 
$|\eta|>8.7$).
Only events with an interaction vertex reconstructed from ITS+TPC tracks with
$|z_{\rm vertex}|<10$~cm were considered in the analysis.

Collisions were classified in centrality classes based on the sum of the 
signal amplitudes in the two V0 scintillator arrays.
Each class is defined in terms of percentiles of the hadronic $\PbPb$ 
cross section, as determined from a fit to the V0 signal amplitude 
distribution based on the Glauber-model description of the geometry 
of the nuclear collision~\cite{Glauber:1955qq,Glauber:1970jm,Miller:2007ri}
and a two-component model for particle production~\cite{ALICEcentr}.
The analysis was performed in two centrality classes: 0--10\% and 20--50\%. 
In total, 16.5$\times 10^6$ events, corresponding to an integrated luminosity 
$L_{\rm int} =(21.5\pm0.7)~\mu{\rm b}^{-1}$, were analysed in the 0--10\% 
centrality class, and 13.5$\times 10^6$ events, 
$L_{\rm int}=(5.9\pm0.2)~\mu{\rm b}^{-1}$, in the 20--50\% class.
The average values of the nuclear overlap function $\TAA$ (defined as 
the convolution of the nuclear density profiles of the colliding 
ions~\cite{Miller:2007ri} and proportional to the  number $N_{\rm coll}$ of 
binary nucleon--nucleon collisions occurring in the Pb--Pb collision) 
are reported in Table~\ref{table:Taa} for the 0--10\% and 20--50\% centrality 
classes, together with their systematic uncertainty estimated as described 
in~\cite{ALICEcentr}.

\begin{table}[h]
\begin{center}
\begin{tabular}{|c|c|c|c|c|}
\hline
  Centrality class&  $\langle \TAA \rangle$ (mb$^{-1}$)&$N_{\rm evt}$ & $L_{\rm int}$ ($\mu$b$^{-1}$) \\ 
\hline
0--10\% & 23.44 $\pm$ 0.76 & 16.4$\times 10^6$ & 21.3$\pm$0.7  \\ 
\hline
20--50\% & 5.46 $\pm$ 0.20 & 13.5$\times 10^6$ & 5.8$\pm$0.2 \\
\hline
\end{tabular}
\end{center}
\caption{Average value of the nuclear overlap function, $\langle \TAA \rangle$,
for the considered centrality classes, expressed as percentiles of the 
hadronic Pb--Pb cross section. 
The values were obtained with a Monte Carlo implementation of the Glauber 
model assuming an inelastic nucleon--nucleon cross section of
64 mb~\cite{ALICEcentr}. 
The number of analysed events and the corresponding integrated luminosity in 
each centrality class are also shown.
The uncertainty on the integrated luminosity derives from the uncertainty of the
hadronic $\PbPb$ cross section from the Glauber model~\cite{ALICEcentr}.
}
\label{table:Taa}
\end{table}

\section{$\Ds$ meson reconstruction and selection}
\label{sec:selection}

$\Ds$ mesons and their antiparticles were reconstructed in the decay channel 
$\DstophipitoKKpi$ (and its charge conjugate), whose branching ratio (BR)
 is  (2.24 $\pm$ 0.10)\%~\cite{PDG}. Other $\Ds$ decay channels can give rise 
to the same $\KKpi$ final state, such as $\DstoKzerostarK$ 
and $\Dstofzeropi$, with BR of (2.58 $\pm$ 0.11)\% and (1.14 $\pm$ 0.31)\%, 
respectively~\cite{PDG}. 
However, as explained in Ref.~\cite{ALICEDspp7}, the applied cuts 
for the selection of the $\Ds$ signal candidates strongly reduce contributions 
from these channels, 
and therefore the measured yield is dominated by the $\DstophipitoKKpi$ decays. 
The decay channel through the $\phi$ resonance was chosen because the narrower width of the
$\phi$ invariant-mass peak with respect to $\fzero$ and $\Kzerostar$ provides the best discrimination 
between signal and background. 

The analysis strategy for the extraction of the signal out of a large 
combinatorial background is based on the reconstruction of decay topologies 
with a secondary vertex significantly displaced  from the interaction point. 
The secondary vertex position and its covariance matrix were determined 
from the decay tracks by using the same analytic $\chi^2$ minimization 
method as for the computation of the primary vertex~\cite{aliceDpp7}.
The resolution on the position of the $\Ds$ decay vertex was estimated with 
Monte Carlo simulations and it was found to be about $100~\mum$.
$\Ds$ mesons have a mean proper decay length $c\tau=150 \pm 2~\mum$~\cite{PDG},
which makes it possible to resolve their decay vertices from the primary 
vertex. 
With the current data sample, the signal of $\Ds$ mesons could be extracted in 
three $\pt$ intervals (4--6, 6--8 and 8--12~$\gev/c$) in the 0--10\% 
centrality class and in two $\pt$ intervals (6--8 and 8--12~$\gev/c$) in the 
20--50\% centrality class.

$\Ds$ candidates were defined from triplets of 
tracks with the proper charge sign combination. 
Tracks were selected requiring $|\eta|<0.8$ and $\pt>0.6~(0.4)~\GeV/c$ 
in the 0--10\% (20--50\%) centrality class. In addition, tracks were also 
required to have at least 70 (out of a maximum of 159) associated hits in the 
TPC, a $\chi^2/{\rm ndf}<2$ of the track momentum fit in the TPC and at 
least one associated hit in one of the two SPD layers. 
With these track selection criteria, the acceptance in 
rapidity for D mesons drops steeply to zero for $|y|\gsim 0.5$ at low $\pt$ 
and for $|y|\gsim 0.8$ at $\pt\gsim 5~\gev/c$. 
A $\pt$-dependent fiducial acceptance cut was therefore applied on the D-meson
rapidity, $|y|<y_{\rm fid}(\pt)$, with $y_{\rm fid}(\pt)$ increasing from 
0.5 to 0.8 in $0<\pt<5~\gev/c$ according to a second order polynomial function
and taking a constant value of 0.8 for $\pt>5~\gev/c$. 

$\Ds$ candidates were filtered by applying kinematical cuts and 
geometrical selections on the decay topology,
together with particle identification criteria. 
The selection criteria were tuned in each $\pt$ interval and centrality class
to have a good statistical significance of the signal, while keeping the 
selection efficiency as high as possible.
It was also checked that background fluctuations were not causing
a distortion in the signal line shape by verifying that the $\Ds$-meson mass 
and its resolution were in agreement with the Particle Data Group (PDG) 
world-average value ($1.969~\gev/c^2$~\cite{PDG}) and the Monte Carlo 
simulation results, respectively.
The resulting selection criteria depend on the transverse momentum of the 
candidate and provide a selection efficiency that increases with increasing 
$\pt$.

The main variables used to select the $\Ds$ decay topology were the
decay length ($L$), defined as the distance between the primary and 
secondary vertices, and the cosine of the pointing angle ($\costhetap$), 
which is the angle between the reconstructed $\Ds$ momentum and the line 
connecting the primary and secondary vertices.
Additional selections were applied on the projections of decay length and 
cosine of pointing angle in the transverse plane $xy$ 
($\Lxy$, $\costhetapxy$), in order to exploit the better resolution on 
the track parameters in that plane. 
A further cut was applied on $\Lxy$ divided by its uncertainty ($\Lxy/\sigma_{\Lxy}$). 
The three tracks were also required to have a small distance to the reconstructed decay vertex, by defining the variable
$\sigma_{\rm vertex}$ as the square root of the sum in quadrature of the distances of each track to the secondary vertex.
To further suppress the combinatorial background, the angles  $\theta^*(\pi)$, i.e.\ the angle between 
the pion in the KK$\pi$ rest frame and the KK$\pi$ flight line in the laboratory frame, and 
$\theta^\prime({\rm K})$, i.e.\ the angle between one of the kaons and the pion in the KK rest 
frame, were exploited. The cut values used for $\Ds$ mesons with 4 $<\pt<$ 6 $\gev/c$ in the 0--10\% centrality class
were: $L$, $\Lxy > 500~\mum$, $\Lxy/\sigma_{\Lxy}>7.5$,  
$\costhetap>$ 0.94, $\costhetapxy>$ 0.94, 
$\sigma_{\rm vertex}< 400~\mum$, 
$\cos \theta^*(\pi)> 0.05$ and $|\cos^3 \theta^\prime({\rm K})|< 0.9$. 
Looser selection criteria were used for $\Ds$ selection at higher $\pt$ and 
in more peripheral events, due to the lower 
combinatorial background.

In addition, to select $\Ds$ mesons decaying in the considered $\phi \pi^+$ 
mode,
with $\phitoKK$, candidates were rejected if none of the two pairs of 
opposite-charged tracks had an invariant mass compatible with the PDG world
average for the $\phi$ mass (1.0195 $\gev/c^2$~\cite{PDG}). 
The difference between the reconstructed K$^+$K$^-$ invariant mass and 
world-average $\phi$ mass was required to be less than 4 $\mev/c^2$ 
(a selection that preserves about 70\% of the signal)
for $\Ds$ candidates in the three $\pt$ 
intervals considered in the 0--10\% centrality class, while looser selections 
were used for semi-central events.

Particle identification was used to obtain a further reduction of the 
background.
Compatibility cuts were applied to the difference between the measured 
signals and those expected for a pion or a kaon.
A track was considered compatible with the kaon or pion hypothesis if both
its $\dEdx$ and time-of-flight were within $3\,\sigma$ from the expected 
values.
Tracks without a TOF signal (mostly at low momentum) were 
identified using only the TPC information 
and requiring a $2\,\sigma$ compatibility with the expected $\dEdx$.
Triplets of selected tracks were required to have two tracks compatible with the kaon
hypothesis and one with the pion hypothesis.
In addition, since the decay particle with opposite charge sign has to be a 
kaon, a triplet was rejected if the opposite-sign track was not compatible 
with the kaon hypothesis.
This particle identification strategy preserves about 85$\%$ of the $\Ds$ 
signal.

For each candidate, two values of invariant mass can be computed, corresponding
to the two possible assignments of the kaon and pion mass to the two same-sign
tracks.
Signal candidates with wrong mass assignment to the same-sign tracks
would give rise to a contribution to the invariant-mass distributions that 
could potentially introduce a bias in the measured raw yield of $\Ds$ mesons.
It was verified, both in data and in simulations, 
that this contribution
is reduced to a negligible level by the particle 
identification selection and by the requirement that the
invariant mass of the two tracks identified as kaons is compatible with 
the $\phi$ mass.

\begin{figure}[!htb]
\begin{center}        
\includegraphics[width=\textwidth]{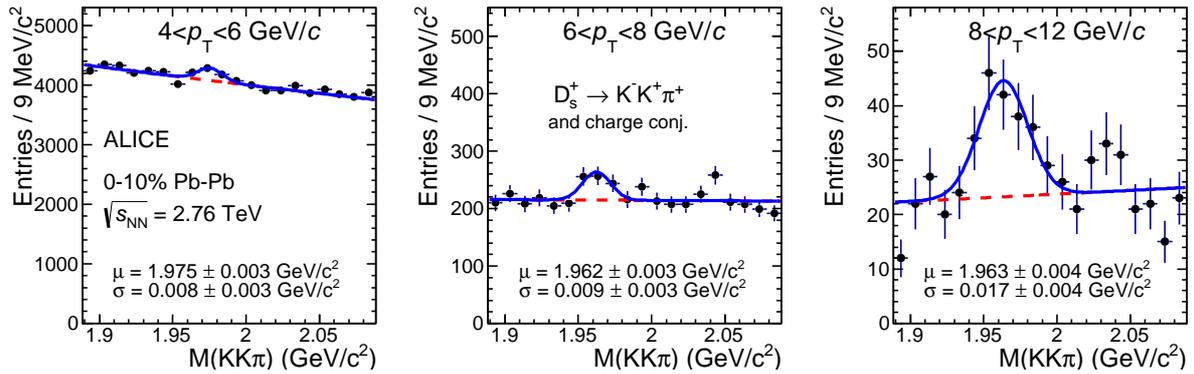}
\caption{Invariant-mass distributions of $\Ds$ candidates and charge
conjugates in the three considered $\pt$ intervals in the 10\% most central Pb--Pb collisions.}
\label{fig:spectra010}
\end{center}
\end{figure}

The invariant-mass distributions of the $\Ds$ candidates (sum of
$\Ds$ and $\Dsminus$ candidates) are shown in Fig.~\ref{fig:spectra010} in 
the three $\pt$ intervals for the 10\% most central $\PbPb$ collisions.
The raw signal yields were extracted by fitting the invariant-mass distributions with a function that consists of the sum 
of a Gaussian term to describe the signal peak and an exponential function to describe the background.
The fit was performed in the invariant-mass range 
$1.88<M({\rm KK\pi})<2.1~\gev/c^2$ in all $\pt$ intervals. 
The lower limit of 1.88 $\gev/c^2$ was chosen to exclude the contribution 
of $\DplustoKKpi$ decays, $\mathrm{BR}=(0.265^{+0.008}_{-0.009})\%$~\cite{PDG}, 
which could give rise to a bump in the background shape for invariant-mass 
values around the $\Dplus$ mass (1.870 $\gev/c^2$)~\cite{PDG}. 
The mean values of the Gaussian functions in all the $\pt$
intervals are compatible within  two times their uncertainty
with the PDG world average for the $\Ds$ mass and 
the Gaussian widths are in agreement with the expected values from 
Monte Carlo simulations.

In Table~\ref{tab:yields} the extracted raw yields of $\Ds$ mesons (sum
of particle and antiparticle), defined as the integral of the Gaussian 
functions, are listed for the different $\pt$ intervals in both the considered
centrality classes, together with the signal-over-background 
(S/B) ratios and the statistical significance (S/$\sqrt{{\rm S} + {\rm B}}$).
The background was evaluated by integrating the background fit functions in
$\pm3\sigma$ around the centroid of the Gaussian.

\begin{table}[!t]
\centering
\vspace{0.5cm}
\begin{tabular}{|c|c|c|c|c|c|} 
\hline \rule{0pt}{2.7ex}
Centrality class & $\pt$ interval &$N^{\rm D_{\rm s}^\pm~raw}$ & S/B (3$\sigma$) & S/$\sqrt{{\rm S} + {\rm B}}$ (3$\sigma$) & $\chi^2$/ndf\\ 
 &(GeV/$c$) & & & & \\ 
\hline \rule{0pt}{2.7ex}
           &\phantom{0}4--6\phantom{0} & 438$\pm$144 & 0.02 & 3.0  & 27.4~/~18\\
0--10\%&\phantom{0}6--8\phantom{0} & 117$\pm$\phantom{0}38 & 0.10 & 3.2 & 17.5~/~18\\
           &\phantom{0}8--12 &  \phantom{0}89$\pm$\phantom{0}21& 0.38 & 5.0 & 26.5~/~18\\
\hline
20--50\%&\phantom{0}6--8\phantom{0} & 197$\pm$\phantom{0}61& 0.07 & 3.5 &\phantom{0}9.9~/~21\\
            &\phantom{0}8--12 &  \phantom{0}52$\pm$\phantom{0}20& 0.29 & 3.4 & 17.9~/~21\\
\hline

\end{tabular}
\caption{Measured raw yields ($N^{\rm D_{\rm s}^\pm~raw}$), signal over background 
(S/B), statistical significance (~S/$\sqrt{{\rm S} + {\rm B}}$~) and $\chi^2$/ndf of the invariant-mass fit for $\Ds$ 
and their antiparticles in the
considered $\pt$ intervals for the 0--10\% and 20--50\% centrality classes. 
The uncertainty on the $\Dspm$ raw yield is the statistical uncertainty 
obtained from the fit.
} 
\label{tab:yields}
\end{table}

\section{Corrections}
\label{sec:corrections}
The raw yields extracted from the fits to the invariant-mass distributions
of $\Ds$ and $\Dsminus$ candidates were corrected to obtain the production 
yields of prompt (i.e.\ not coming from weak decays of B 
mesons) $\Ds$ mesons.
The $\pt$-differential yield of prompt $\Ds$ was computed as
\begin{equation}
  \label{eq:dNdpt}
  \left.\frac{{\rm d} N^{\rm D_{\rm s}^+}}{{\rm d}\pt}\right|_{|y|<0.5}=
  \frac{1}{ \Delta \pt}\frac{1}{{\rm BR} \cdot N_{\rm evt}}\frac{\left.f_{\rm prompt}(\pt)\cdot \frac{1}{2} N^{\rm D_{\rm s}^\pm~raw}(\pt)\right|_{|y|<y_{\rm fid}}}{ 2 y_{\rm fid}(\pt) \,({\rm Acc}\times\epsilon)_{\rm prompt}(\pt)}\,,
\end{equation}
where $N^{\rm D_{\rm s}^\pm~raw}(\pt)$ are the values of the raw yields 
(sum of particles and antiparticles) reported in Table~\ref{tab:yields},
 which were corrected for the B-meson decay feed-down contribution 
(i.e.\ multiplied by the prompt fraction $f_{\rm{prompt}}$), divided by the 
acceptance-times-efficiency for prompt $\Ds$ mesons, 
$(\rm Acc \times \epsilon)_{\rm{prompt}}$, and divided by a factor of two to obtain the 
charge (particle and antiparticle) averaged yields.
The corrected yields were divided by the decay channel branching ratio (BR), 
the $\pt$ interval width ($\Delta \pt$), the rapidity coverage 
($2 y_{\rm fid}$) and the number of analysed events ($N_{\rm evt}$).  

The correction for the acceptance and the efficiency was determined using 
Monte Carlo simulations. \mbox{Pb--Pb} collisions at 
$\sqrtsNN =2.76~\tev$ were simulated using the HIJING~v1.383
event generator~\cite{Hijing}.
Prompt and feed-down $\Ds$ (and $\Dsminus$) signals were 
added with the PYTHIA~v6.4.21 generator~\cite{PYTHIA}.
In order to minimize the bias on the detector occupancy,
the number of D mesons injected into each
HIJING event was adjusted according to the Pb--Pb collision centrality.
The $\pt$ distribution of the generated $\Ds$ mesons in the 
0--10\% centrality class was weighted in order to match the shape measured 
for $\Dzero$ mesons in central Pb--Pb collisions~\cite{Adam:2015sza}.
For the 20--50\% centrality class, the generated $\pt$ distribution was 
defined based on FONLL perturbative QCD calculations~\cite{FONLL,FONLL2012} 
multiplied by the nuclear modification factor predicted by the BAMPS partonic 
transport model~\cite{BAMPS}, which reproduces the measured non-strange 
D-meson $\RAA$ in semi-central collisions within uncertainties~\cite{ALICEDv2artic}.

The generated particles were transported through the ALICE detector using
the GEANT3~\cite{Geant3} particle transport package together with a detailed 
description of the geometry of the apparatus and of the detector response.
The simulation was tuned to reproduce the position and width of the 
interaction vertex distribution, the number of active electronic channels and 
the accuracy of the detector calibration, and their time evolution within 
the Pb--Pb data taking period. 

\begin{figure}[t!]
\begin{center}

\includegraphics[width=0.48\textwidth, height=7.2cm]{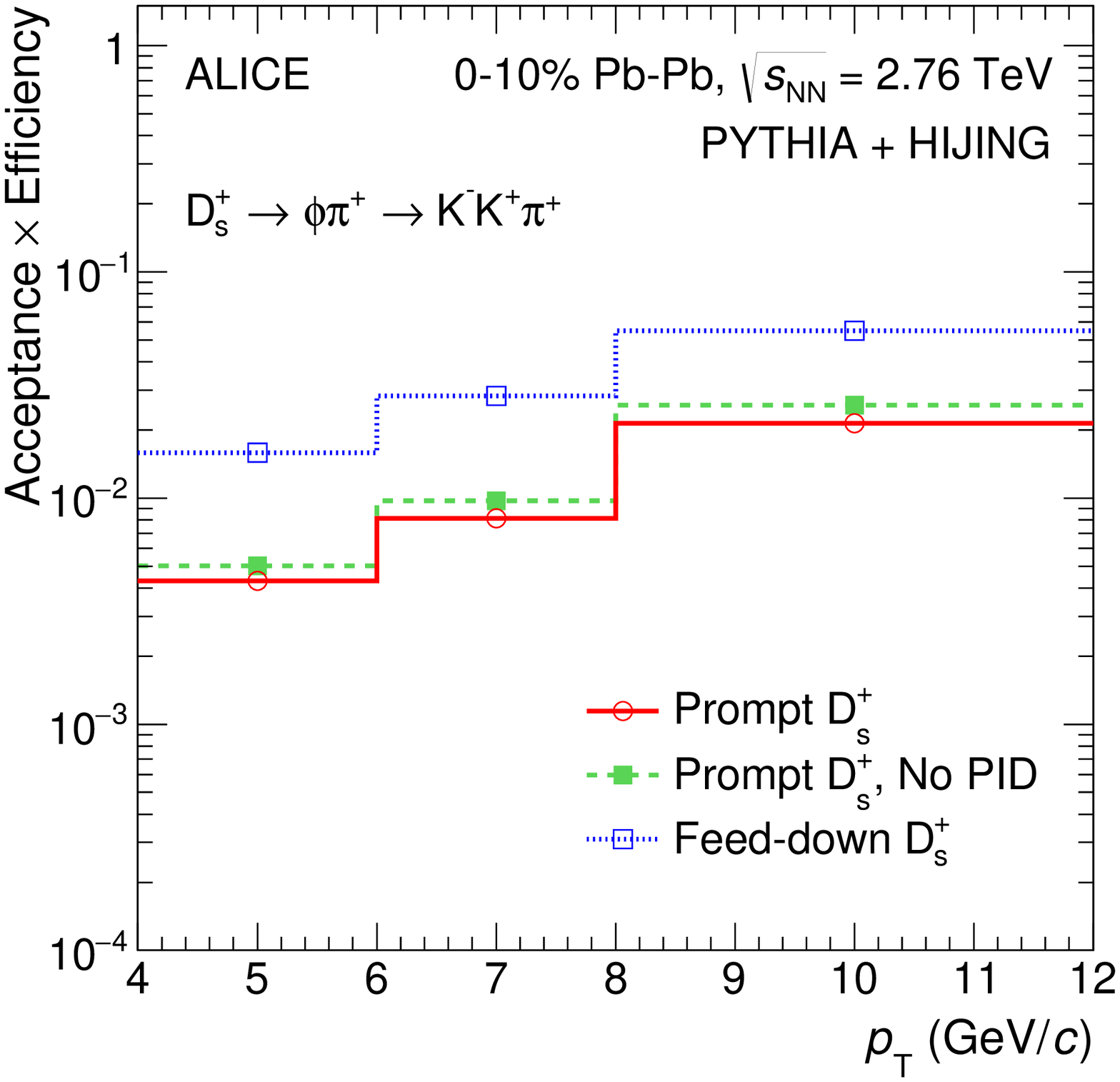}
\includegraphics[width=0.48\textwidth, height=7.2cm]{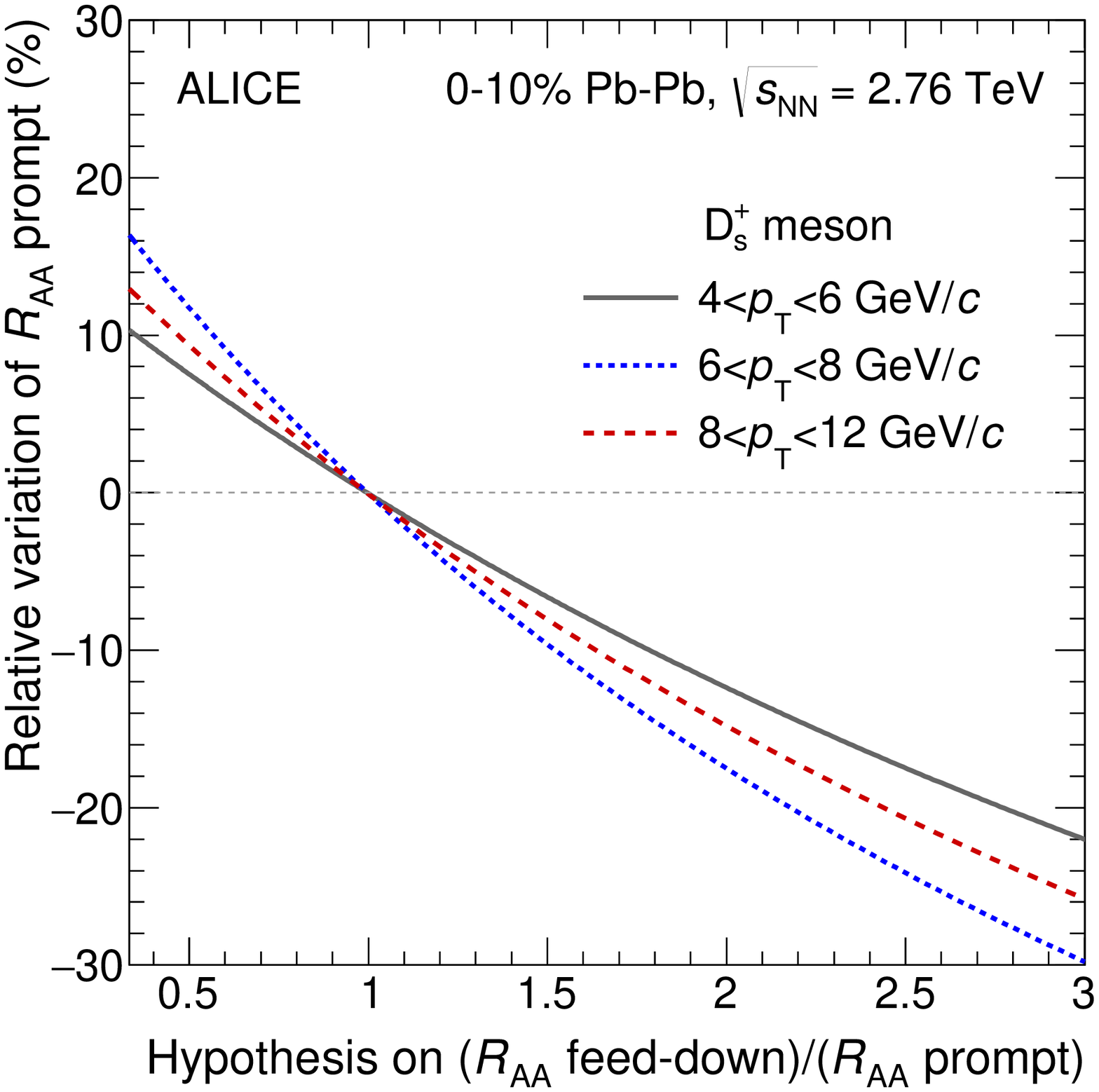}

\end{center}
\caption{Left: Acceptance-times-efficiency for $\Ds$ mesons in the 10\% most central
Pb--Pb collisions.
The efficiencies for prompt (solid lines) and feed-down (dotted lines)
$\Ds$ mesons are shown. Also displayed, for comparison, the efficiency for 
prompt $\Ds$ mesons without PID selections (dashed lines). Right: Relative
  variation  of the prompt $\Ds$-meson yield in the 0--10\%
centrality class as a function of the hypothesis on  $\RAA^{\textnormal{feed-down}}/\RAA^{\rm prompt}$ for the B 
feed-down subtraction approach based on Eq.~(\ref{eq:fpromptNbMethod}).}
\label{fig:effic}
\end{figure}

The efficiencies were evaluated in centrality classes corresponding
to those used in the analysis of the data in terms of charged-particle 
multiplicity, hence of detector occupancy.
In the left-hand panel of Fig.~\ref{fig:effic}, the $({\rm Acc}\times\epsilon)$ values for prompt 
and feed-down $\Ds$ mesons with rapidity $|y|<y_{\rm fid}$ are shown for the 
 0--10\% centrality class. 
The same figure shows also the $({\rm Acc}\times\epsilon)$ 
values for the case without the PID selections, demonstrating that 
this selection is about 85\% efficient for the signal.

The magnitude of $({\rm Acc}\times\epsilon)$ increases with increasing 
$\pt$, from 0.4\% in the lowest $\pt$ interval up to 2\%
in $8<\pt<12~\gev/c$.
The $({\rm Acc}\times\epsilon)$ values for $\Ds$ from beauty-hadron 
decays are larger than those for prompt $\Ds$ by a factor of 
approximately  2.5--3.5 depending on $\pt$,
because the decay vertices of the feed-down $\Ds$ mesons 
are more displaced from the primary vertex and they are, therefore, more 
efficiently selected by the analysis cuts. 
The efficiency of the selections used in the centrality interval 20--50\%
is higher by a factor of about two with respect to that in the most central 
events, because the smaller combinatorial background in semi-peripheral 
collisions allowed the usage of looser selections on the $\Ds$ candidates.

The ratio of prompt to inclusive contributions in the $\Ds$-meson raw yield, 
$f_{\rm prompt}$, was evaluated using a procedure similar to the one 
adopted for the pp measurement~\cite{ALICEDspp7}.
The contribution of feed-down from B decays in the raw yield depends on 
$\pt$ and on the applied geometrical selection criteria. 
The feed-down contribution was estimated using 
the beauty-hadron production cross section from FONLL perturbative QCD
calculations for pp collisions at $\sqrt{s}=2.76~\tev$ scaled by the average 
nuclear overlap function $\langle \TAA \rangle$ in each centrality class,
the B$\rightarrow$D+$X$ decay kinematics from the EvtGen package~\cite{Evtgen}
and the Monte Carlo efficiencies for feed-down $\Ds$ mesons.
The resulting sample of feed-down $\Ds$ mesons is composed of two 
contributions: about 50\% of the feed-down originates from 
B$^0_{\rm s}$-meson decays, while the remaining 50\% comes from decays of 
non-strange B mesons (B$^0$ and B$^+$).
A hypothesis on the nuclear 
modification factor of feed-down $\Ds$ mesons, $\RAA^{\textnormal{feed-down}}$, was 
introduced to account for the different modification of beauty and charm 
production in Pb--Pb collisions and for the possible enhancement
of the B$^{0}_{\rm s}$ over non-strange B-meson yield due to the effect of 
hadronisation via recombination~\cite{TAMULHC}.
The fraction of prompt $\Ds$ yield was therefore computed in each $\pt$ 
interval as
\begin{equation}
  \label{eq:fpromptNbMethod}
\begin{aligned}
f_{\rm prompt} &= 1-\frac{N^{\rm D_s^+~\textnormal{feed-down}~raw}}{N^{\rm D_s^+~raw}}=\\
 &= 1 - \langle \TAA \rangle 
	 \cdot  \left( \frac{{\rm d}^2 \sigma}{{\rm d}y \, {\rm d}\pt } \right)^{{\sf FONLL}} _{\textnormal{feed-down}} \cdot 
	 & \RAA^{\textnormal{feed-down}} \cdot \frac{({\rm Acc}\times\epsilon)_{\textnormal{feed-down}}\cdot 2y_{\rm fid} \, \Delta\pt \cdot {\rm BR} \cdot N_{\rm evt}  }{ N^{\rm D_s^\pm~raw }  / 2} \, ,
\end{aligned}
\end{equation}
where $({\rm Acc}\times\epsilon)_{\textnormal{feed-down}}$ is the 
acceptance-times-efficiency for feed-down $\Ds$ mesons.
To determine the central value of $f_{\rm prompt}$, it was assumed that the 
nuclear modification factors of feed-down and prompt $\Ds$ mesons were equal 
($\RAA^{\textnormal{feed-down}}=\RAA^{\rm prompt}$). 
The resulting feed-down contribution is about 20--25\% depending on the
$\pt$ interval.
To determine the systematic uncertainty the hypothesis
was varied in the range $1/3<\RAA^{\textnormal{feed-down}}/\RAA^{\rm prompt}<3$, as 
discussed in detail in Section~\ref{sec:systematics}.
It should be noted that the central value and the range of the hypothesis
on $\RAA^{\textnormal{feed-down}}/\RAA^{\rm prompt}$ differ from those used for
non-strange D mesons in Refs.~\cite{ALICEDv2lett,ALICEDv2artic,Adam:2015sza},
owing to the unknown role of recombination in the beauty sector, which
could enhance the ratio of B$^0_{\rm s}$ over non-strange B mesons, and to the
large fraction of feed-down $\Ds$ mesons originating from non-strange B-meson 
decays.

The nuclear modification factor of $\Ds$ mesons was computed as
\begin{equation}
R_{\rm AA}(\pt)=\frac{{\rm d} N^{\rm D_{\rm s}^+}_{\rm AA}/{\rm d}\pt}{ \av{T_{\rm AA}} {\rm d}\sigma^{\rm D_{\rm s}^+}_{\rm pp}/{\rm d}\pt}.
\end{equation} 
The values of the average nuclear overlap function, $\av{T_{\rm AA}}$, for the considered centrality 
classes are reported in Table~\ref{table:Taa}. 
The $\pt$-differential cross section of prompt $\Ds$ mesons with $|y| < 0.5$ 
in pp collisions at $\sqrt{s}=2.76~\TeV$, used as reference for $\RAA$, 
was obtained by scaling in energy the measurement at 
$\sqrt{s}=7~\TeV$~\cite{ALICEDspp7}.
The ratio of the cross sections from 
FONLL pQCD calculations~\cite{FONLL2012} at $\sqrt{s}=2.76$ and $7~\tev$ was used
as the scaling factor.
Since FONLL does not have a specific prediction for $\Ds$ mesons,
the cross sections of the D-meson admixture (70\% of $\Dzero$ and 30\% 
of $\Dplus$) were used for the scaling.
The theoretical uncertainty on the scaling factor was evaluated by 
considering the envelope of the results obtained
by varying independently the factorisation and renormalisation scales and the 
charm quark mass, as explained in detail in Ref.~\cite{EnScaling}.
For $\Dzero$, $\Dplus$ and $\Dstar$ mesons, the result of the scaling was 
validated by comparison with data~\cite{Dpp2.76}.

\section{Systematic uncertainties}
\label{sec:systematics}
The systematic uncertainties on the prompt $\Ds$-meson yields in Pb--Pb 
collisions are summarised in Table~\ref{tab:Syst}.

\begin{table}[!tb]
\centering
\begin{tabular}{|l||c|c|c||c|c||} 
\hline 
 & \multicolumn{3}{c||}{0--10\% centrality}
 & \multicolumn{2}{c||}{20--50\% centrality}\\[1ex]
\cline{2-6}
 & \multicolumn{3}{c||}{$\pt$ interval ($\gev/c$)}
 & \multicolumn{2}{c||}{$\pt$ interval ($\gev/c$)}\\
 & 4--6 & 6--8 & 8--12 & 6--8 & 8--12\\
\hline
Raw yield extraction  & \phantom{0}8\% & \phantom{0}8\% & \phantom{0}8\% & \phantom{0}8\% & \phantom{0}8\%\\
Tracking efficiency   & 15\% & 15\% & 15\% & 15\% & 15\%\\
Selection efficiency  & 20\% & 20\% & 20\% & 20\% & 20\%\\
PID efficiency        & \phantom{0}7\% & \phantom{0}7\% & \phantom{0}7\% & \phantom{0}7\% & \phantom{0}7\%\\
MC $\pt$ shape        & \phantom{0}2\% & \phantom{0}1\% & \phantom{0}1\% & \phantom{0}1\% &  \phantom{0}1\%  \\
Feed-down from B   & & & & &\\
$\qquad$   FONLL feed-down corr.   & $_{-28} ^{+\phantom{0}6}\%$ & $_{-27} ^{+10}\%$
                           & $_{-27} ^{+\phantom{0}7}\%$ & $_{-20} ^{+\phantom{0}6}\%$
                           & $_{-25} ^{+\phantom{0}7}\%$\\[1ex]
$\qquad$   $\RAA^{\textnormal{feed-down}}/\RAA^{\rm prompt}$ (Eq.~(\ref{eq:fpromptNbMethod}))
                           & $_{-22} ^{+10}\%$ & $_{-30} ^{+16}\%$
                           & $_{-26} ^{+13}\%$ & $_{-22} ^{+11}\%$
                           & $_{-24} ^{+12}\%$\\[1ex]

\hline
Centrality limits     & \multicolumn{3}{c||}{$<1$\%}
                      & \multicolumn{2}{c||}{$<1$\%}\\
\hline
Branching ratio       & \multicolumn{5}{c||}{4.5\%} \\
\hline
\end{tabular}
\caption{Relative systematic uncertainties on $\pt$-differential yields
of prompt $\Ds$ mesons in Pb--Pb collisions for the two considered centrality classes.}
\label{tab:Syst}
\end{table}

The systematic uncertainty on the raw yield extraction was estimated from the 
distribution of the results obtained by repeating the fit to the
invariant-mass spectra varying i) the fit range and ii) the probability 
distribution functions used to model the signal and background contributions.
In particular, a second order polynomial function was used as an alternative
functional form to describe the background.
The signal line shape was varied by using Gaussian functions with
mean and width fixed to the world-average $\Ds$ mass and to the values 
expected from Monte Carlo simulations, respectively.
Furthermore, the raw yield was also extracted by counting the entries in the 
invariant-mass distributions after subtraction of the background estimated 
from a fit to the side bands of the $\Ds$ peak.
In case of fitting in an extended mass range, it was verified that
the effect on the $\Ds$ yield due to the possible bump produced 
in the candidate invariant-mass distribution by 
$\Dplus \to \phi \pi^+ \to \KKpi$ decays was negligible.
An additional test was performed by fitting the $\Ds$ candidate invariant-mass 
distribution after subtracting the background estimated by coupling a 
pion track with K$^+$K$^-$ pairs having an invariant mass in the side bands of 
the $\phi$ peak. 
The uncertainty was estimated to be 8\% in all $\pt$ intervals.

The contribution to the measured yield from $\Ds$ decaying into the $\KKpi$ 
final state via other resonant channels (i.e.\ not via a $\phi$ meson) 
was found to be negligible, due to the much 
lower selection efficiency, as discussed in Ref.~\cite{ALICEDspp7}.

Other contributions to the systematic uncertainty originate from the imperfect 
implementation of the detector description in the Monte Carlo simulations,
which could affect the particle reconstruction, the $\Ds$ selection 
efficiency, and the kaon and pion identification.

The systematic uncertainty on the tracking efficiency (including the effect 
of the track selection) was estimated by comparing  
the efficiency (i) of track finding in the TPC and (ii) of track prolongation 
from the TPC to the ITS between data and simulations, and (iii) by varying 
the track quality selections.
The estimated uncertainty is 5\% per track, which results in 15\% for the
three-body decay of $\Ds$ mesons.

The effect of residual discrepancies between data and simulations 
on the variables used to select the $\Ds$ candidates was estimated by 
repeating the analysis with different 
geometrical selections on the decay topology and varying the
cut on the compatibility between the K$^+$K$^-$ invariant mass and the 
$\phi$ mass.
A systematic uncertainty  of 20\% was estimated from the spread
of the resulting corrected yields. 

The systematic uncertainty induced by a different efficiency for particle 
identification in data and simulations was estimated by comparing the
corrected $\Ds$ yields obtained using different PID approaches,
testing both looser and tighter cuts with respect to the 
baseline selection described in Section~\ref{sec:corrections}.
Due to the limited statistical significance, an analysis without PID selection
could not be carried out. 
Such a test was performed in the analysis of $\Dzero$ ($\to {\rm K}^-\pi^+$), $\Dplus$ ($\to {\rm K}^-\pi^+\pi^+$) and 
$\Dstar$ ($\to {\rm D}^0 \pi^+$) and a 5\% uncertainty was estimated for the case 
of 3$\,\sigma$ cuts on $\dEdx$ and time-of-flight signals, which
correspond to the loosest selections that could be tested for the $\Ds$.
Based on all these checks a systematic uncertainty of 7\% on the PID
selection efficiency was estimated.

The efficiency is also sensitive to differences between the real 
and simulated $\Ds$ momentum distributions.
The effect depends on the width of the 
$\pt$ intervals and on the variation of the efficiency within them. 
A systematic uncertainty was defined from the relative difference 
among the efficiencies obtained using different $\pt$ shapes for
the generated $\Ds$ mesons, namely the measured d$N$/d$\pt$ of
$\Dzero$ mesons in central Pb--Pb collisions, the $\pt$ shape predicted
by FONLL pQCD calculations with and without the nuclear modification 
predicted by the BAMPS partonic transport model.
The resulting contribution to the systematic uncertainty was found to be
2\% for the momentum interval $4<\pt<6~\gev/c$, where the selection efficiency
is strongly $\pt$ dependent, and 1\% at higher $\pt$.

The systematic uncertainty due to the subtraction of $\Ds$ mesons from 
B-meson decays was estimated following the procedure described 
in Ref.~\cite{ALICEDRAA}.
The contribution of the uncertainties inherent in the FONLL perturbative 
calculation was included by varying the heavy-quark masses and the 
factorisation and renormalisation scales, $\mu_{\rm F}$ and $\mu_{\rm R}$, 
independently in the ranges $0.5<\mu_{\rm F}/\mt<2$, $0.5<\mu_{\rm R}/\mt<2$, 
with the constraint $0.5<\mu_{\rm F}/\mu_{\rm R}<2$, 
where $\mt=\sqrt{\pt^2+m_{\rm Q}^2}$.
Furthermore, the prompt fraction obtained in each $\pt$ interval was compared 
with the results of a different procedure in which the FONLL
cross sections for prompt and feed-down D mesons and their respective 
Monte Carlo efficiencies were the input for evaluating the correction factor
\begin{equation}
f^\prime_{\rm prompt} = \left( 	1 + 	\frac{({\rm Acc}\times\epsilon)_{\textnormal{feed-down}}}{({\rm Acc}\times\epsilon)_{{\rm prompt}}}	\cdot
		 \frac{ \left(\frac{{\rm d}^2 \sigma}{{\rm d}y \, {\rm d} \pt } \right)^{{\sf FONLL}}_{\textnormal{feed-down}} }{ \left(\frac{{\rm d}^2 \sigma}{{\rm d}y \, {\rm d} \pt } \right)^{{\sf FONLL}}_{ {\rm prompt} } } 
		 \cdot \frac{ \RAA^{\textnormal{feed-down}} } { \RAA^{\rm prompt} }
\right)^{-1} \, .
\label{eq:fpromptfcMethod}
\end{equation}
 
Since FONLL does not have a specific prediction for $\Ds$ mesons, 
four different approaches were used to compute the predicted $\pt$
shapes of promptly produced $\Ds$, 
$\left( {\rm d}^2 \sigma/{\rm d}y \, {\rm d} \pt \right)^{{\sf FONLL}}_{ {\rm prompt} }$, 
as explained in detail in Ref.~\cite{ALICEDspp7}: 
(i) FONLL prediction for the admixture of charm hadrons; 
(ii) FONLL prediction for $\Dstar$ mesons (the $\Dstar$ mass being close to that of the $\Ds$); 
(iii) FONLL prediction for c quarks and fragmentation functions 
from~\cite{Braaten} with parameter $r=(m_{\rm D}-m_{\rm c})/m_{\rm D}$ ($m_{\rm D}$ and $m_{\rm c}$ being the masses of the considered D-meson species and of the 
c quark, respectively);
(iv) FONLL prediction for c quarks and fragmentation functions 
from~\cite{Braaten} with parameter $r=0.1$ (as used in FONLL calculations) 
for all meson species.
In the latter two cases, the ${\rm D}_{\rm s}^{*+}$ mesons produced in the c 
quark fragmentation were made to decay with PYTHIA and the resulting 
$\Ds$ were summed to the primary ones to obtain the prompt yield.
The systematic uncertainty due to the B feed-down subtraction was finally
evaluated as the envelope of the results obtained with the two methods,
namely Eq.~(\ref{eq:fpromptNbMethod}) and (\ref{eq:fpromptfcMethod}),
when varying the FONLL parameters and the c$\rightarrow \Ds$ fragmentation 
function used to determine $\left( {\rm d}^2 \sigma/{\rm d}y \, {\rm d} \pt \right)^{{\sf FONLL}}_{ {\rm prompt} }$ in Eq.~(\ref{eq:fpromptfcMethod}).

The contribution due to the different nuclear modification factor of prompt and
feed-down $\Ds$ mesons was estimated by varying the hypothesis on 
$\RAA^{\textnormal{feed-down}}/\RAA^{\rm prompt}$ in the range $1/3<\RAA^{\textnormal{feed-down}}/\RAA^{\rm prompt}<3$ for both feed-down subtraction methods.
The variation of the hypothesis is motivated by the combined effect on the 
$\RAA$ of (i) the different energy loss of charm and beauty quarks in the QGP, 
as predicted by energy loss models and supported by experimental data on D 
meson and non-prompt $\rm J/\psi$ $\RAA$ at the 
LHC~\cite{ALICEDRAA,CMSJpsi,ALICEnonpromptJpsi,Adam:2015sza,ALICEDRaaNpart};
(ii) the possibly different contribution of coalescence
in charm and beauty quark hadronisation, leading to a different 
abundance of $\Ds$ and B$_{\rm s}^0$ mesons relative to non-strange mesons; and
(iii) the possibly different modulation of D and B spectra due to radial flow.
The resulting uncertainty for the case of B feed-down subtraction approach 
based on Eq.~(\ref{eq:fpromptNbMethod}) is shown in the right-hand panel of 
Fig.~\ref{fig:effic} for the three $\pt$ intervals in the 0--10\% centrality 
class.

The Pb--Pb data are also affected by a systematic uncertainty 
on the determination of the limits of the centrality classes, due to the 
1.1\% relative uncertainty on the fraction of the total hadronic cross section 
used in the Glauber fit~\cite{ALICEcentr}.
This contribution was estimated from the variation of the D-meson 
d$N$/d$\pt$ when the limits of the centrality classes are shifted by 
$\pm$1.1\%. 
The resulting uncertainty, which is common to all $\pt$ bins, is less 
than 1\% for both the 0--10\% and the 20--50\% centrality classes. 

Finally, the 4.5\% uncertainty on the branching ratio~\cite{PDG}
was considered.

In the calculation of the $\RAA$, the uncertainties on the reference cross 
section for pp collisions, the Pb--Pb yields, and the average 
nuclear overlap function were considered.

For the pp reference, the uncertainties on the measurement 
at $\sqrt{s}=7~\TeV$, described in Ref.~\cite{ALICEDspp7} and those due to
the FONLL-based scaling to $\sqrt{s}=2.76~\TeV$, described in 
Section~\ref{sec:corrections}, were summed in quadrature.
The contributions to the systematic uncertainty
on the pp reference cross section are reported in Table~\ref{tab:Systpp}.

\begin{table}[!tb]
\centering
\begin{tabular}{|l|c|c|c|} 
\hline 
 & \multicolumn{3}{c|}{$\pt$ interval ($\gev/c$)}\\
 & 4--6 & 6--8 & 8--12 \\
\hline
Data systematics in pp & 26\% & 25\% & 29\%\\[1ex] 
Feed-down from B      & $_{-17} ^{+\phantom{0}4}\%$
                      & $_{-15} ^{+\phantom{0}6}\%$ 
                      & $_{-17} ^{+\phantom{0}5}\%$\\[2ex]
$\sqrt{s}$-scaling of the pp reference & $_{-\phantom{0}7} ^{+14}\%$
                                       & $_{-\phantom{0}6} ^{+10}\%$
                                       & $_{-\phantom{0}5} ^{+\phantom{0}8}\%$\\[1ex]
\hline
Normalisation         & \multicolumn{3}{c|}{3.5\%}\\
Branching ratio       & \multicolumn{3}{c|}{4.5\%} \\
\hline
\end{tabular}
\caption{Relative systematic uncertainties on the pp reference cross section.
The row labeled `Data systematics' reports the sum in quadrature of the
contributions due to raw yield extraction, tracking efficiency, selection 
efficiency, PID efficiency, MC $\pt$ shape and 'other resonant channels' 
from Ref.~\cite{ALICEDspp7}.}
\label{tab:Systpp}
\end{table}

The uncertainties on the pp reference were added in quadrature to those on the
Pb--Pb prompt $\Ds$ yields, described above, except for the BR that cancels out in the ratio
and the feed-down contribution deriving from FONLL uncertainties,
that partly cancels in the ratio.
This contribution was evaluated by comparing the $\RAA$ values obtained
with the two methods for feed-down correction of 
Eq.~(\ref{eq:fpromptNbMethod}) and (\ref{eq:fpromptfcMethod}) and
with the different heavy quark masses, fragmentation functions, factorisation 
and renormalisation scales used in FONLL.
In this study, these variations were done simultaneously for the Pb--Pb 
yield and for the pp reference cross section, so as to take into account the 
correlations of these sources in the numerator and denominator of $\RAA$.

Finally, the $\RAA$ normalisation uncertainty was computed as the quadratic 
sum of the 3.5\% pp normalisation uncertainty~\cite{ALICEDspp7}, the 
contribution due to the 1.1\% uncertainty on the fraction of hadronic cross 
section used in the Glauber fit discussed above, and the uncertainty 
on $\av{\TAA}$, which is of 3.2\% and 3.7\% for the 0--10\% and 20--50\% 
centrality classes, respectively.

\section{Results}
\label{sec:results}
\begin{figure}[t!]
\begin{center}
\includegraphics[width=0.48\textwidth]{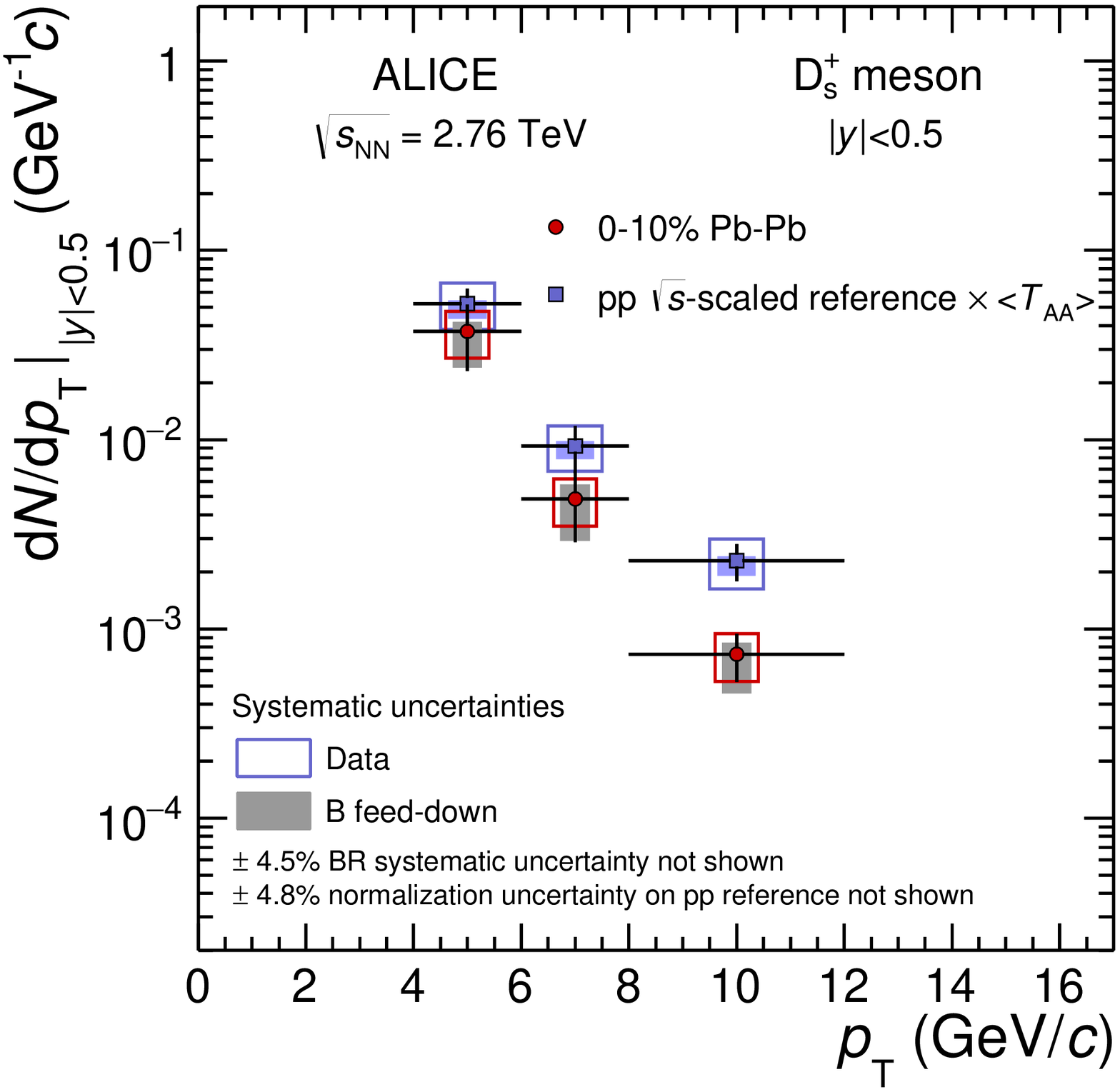}
\includegraphics[width=0.48\textwidth]{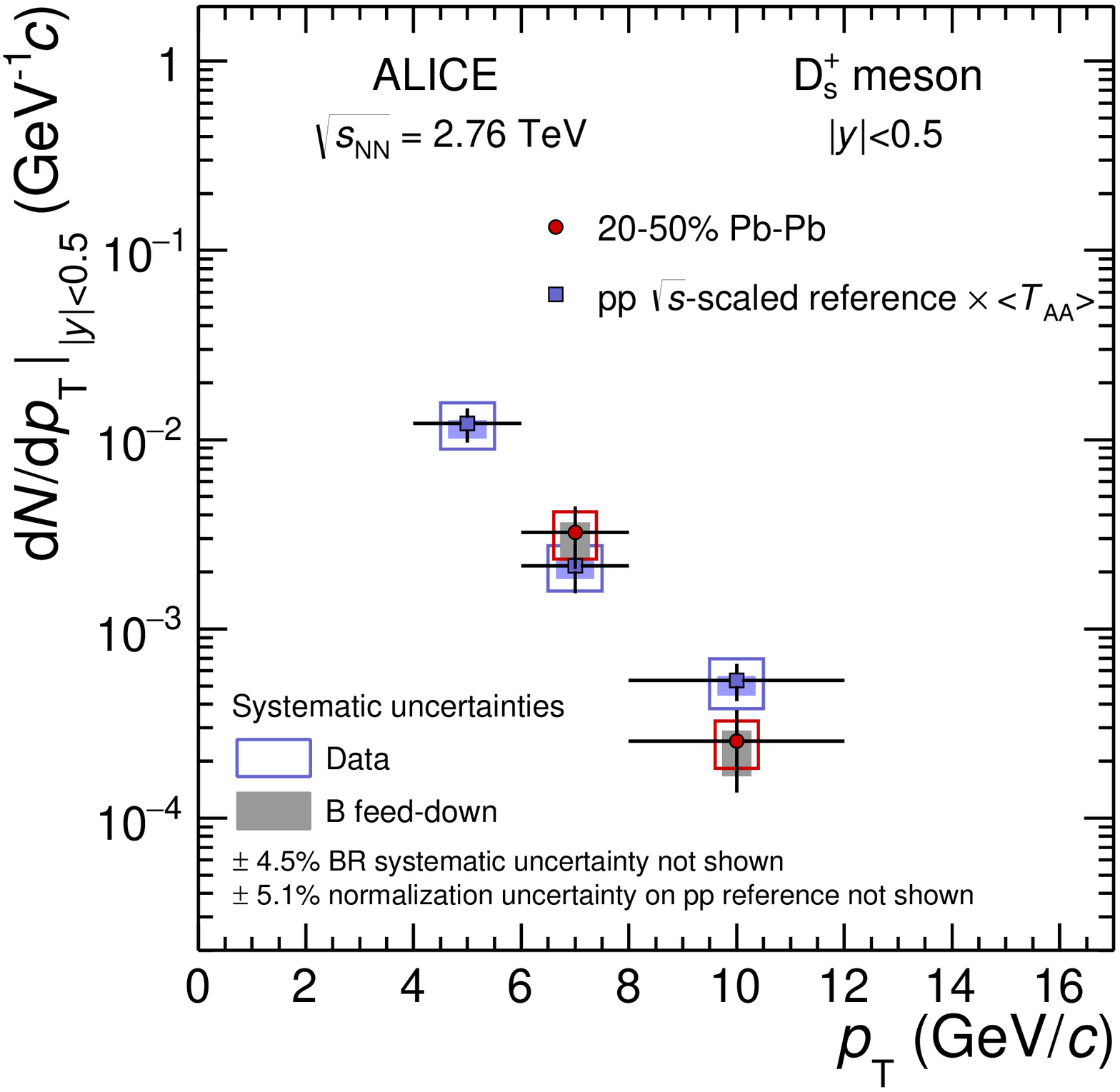}
\caption{Transverse momentum distributions $\mathrm{d} N/\mathrm{d}\pt$ of 
prompt $\Ds$ mesons in the 0--10\% (left panel) and 20--50\% (right panel)
centrality classes in Pb--Pb collisions at $\sqrtsNN=2.76~\tev$.
Statistical uncertainties (bars), systematic uncertainties from data 
analysis (empty boxes) and systematic uncertainties due to beauty feed-down 
subtraction (shaded boxes) are shown. 
The reference pp distributions $\av{\TAA}\,\mathrm{d} \sigma/\mathrm{d}\pt$ 
are shown as well. }
\label{fig:dndpt}
\end{center}
\end{figure}

The transverse momentum distributions d$N$/d$\pt$ of prompt
$\Ds$ mesons in Pb--Pb collisions are shown in Fig.~\ref{fig:dndpt}, for the 0--10\% and 
20--50\% centrality classes.
The yields reported in Fig.~\ref{fig:dndpt} refer to particles only, 
since they were computed as the average of particles and antiparticles under
the assumption that the production cross section is the same for 
${\rm D_s^+}$ and ${\rm D_s^-}$.
The vertical error bars represent the statistical uncertainties.
The symbols are positioned horizontally at the centre of each $\pt$ interval,
with the horizontal bars representing the width of the $\pt$ interval. 
The systematic uncertainties from data analysis are shown as empty 
boxes around the data points, while those due to the B feed-down subtraction,
which include the contributions of the FONLL uncertainties and 
of the variation of the hypothesis on 
$R_{{\rm AA}}^{\textnormal{feed-down}}/R_{{\rm AA}}^{\rm prompt} $, are displayed as shaded 
boxes.
The normalisation uncertainties are reported as text on the figures.

The $\pt$-differential yields measured in Pb--Pb collisions are compared
to the reference yields in pp collisions at the same energy, scaled by the 
nuclear overlap function $\av{\TAA}$, reported in Table~\ref{table:Taa}. 
The pp reference at $\sqrt{s}=2.76~\TeV$ is obtained by scaling the
cross section measured at 7~TeV as described in Section~\ref{sec:corrections}.
A clear suppression of the $\Ds$-meson yield in the 10\% most central Pb--Pb 
collisions relative to the binary-scaled pp yields is observed in the 
highest $\pt$ interval ($8<\pt<12~\gev/c$).
In the 20--50\% centrality class, an indication of suppression is found in 
$8<\pt<12~\gev/c$.
At lower $\pt$, in both centrality classes, it is not possible to conclude on the presence of 
a suppression of the $\Ds$-meson yield in heavy-ion collisions with respect to the
pp reference.

\begin{figure}[t!]
\begin{center}
\includegraphics[width=0.48\textwidth]{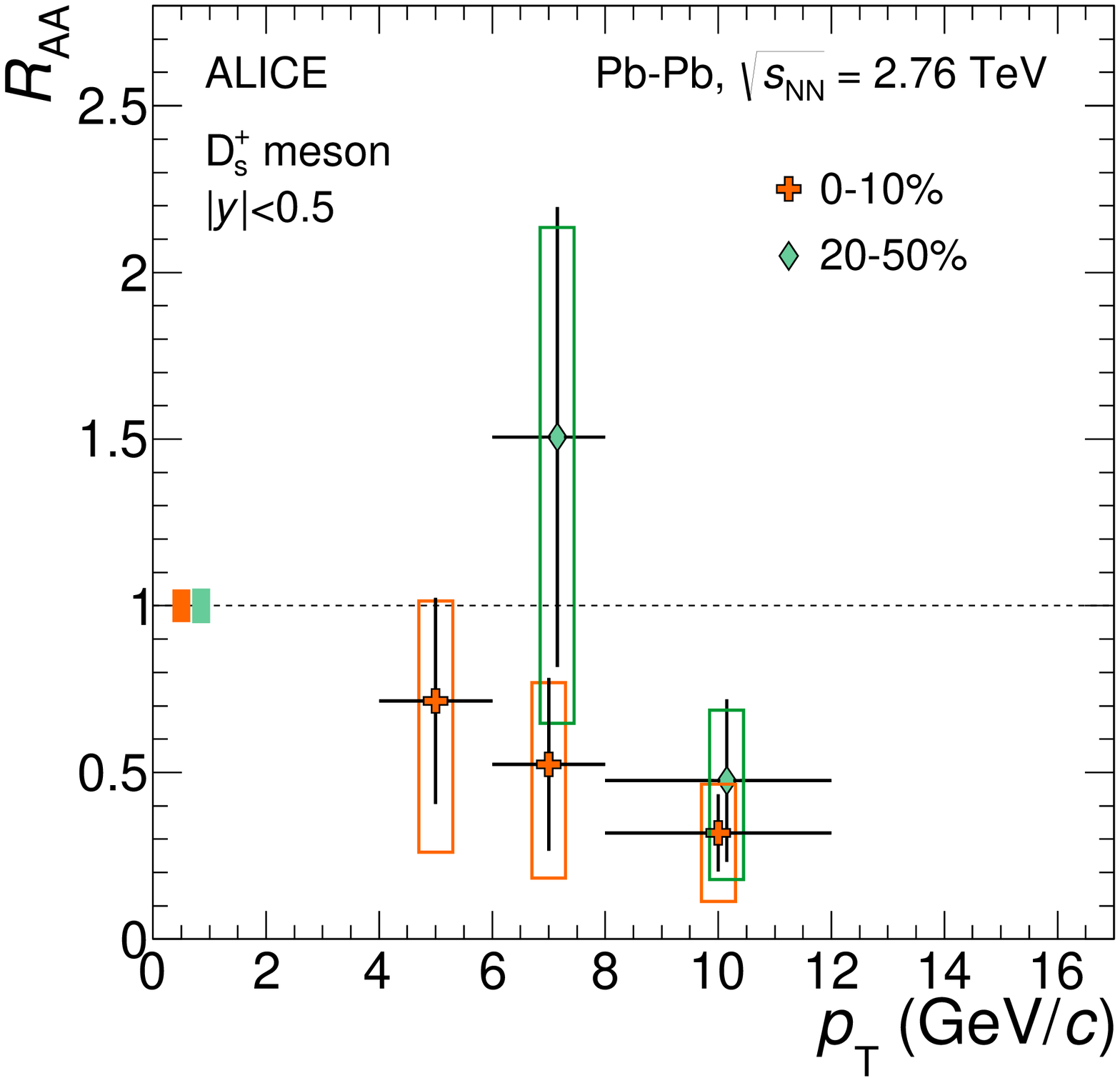}
\includegraphics[width=0.48\textwidth]{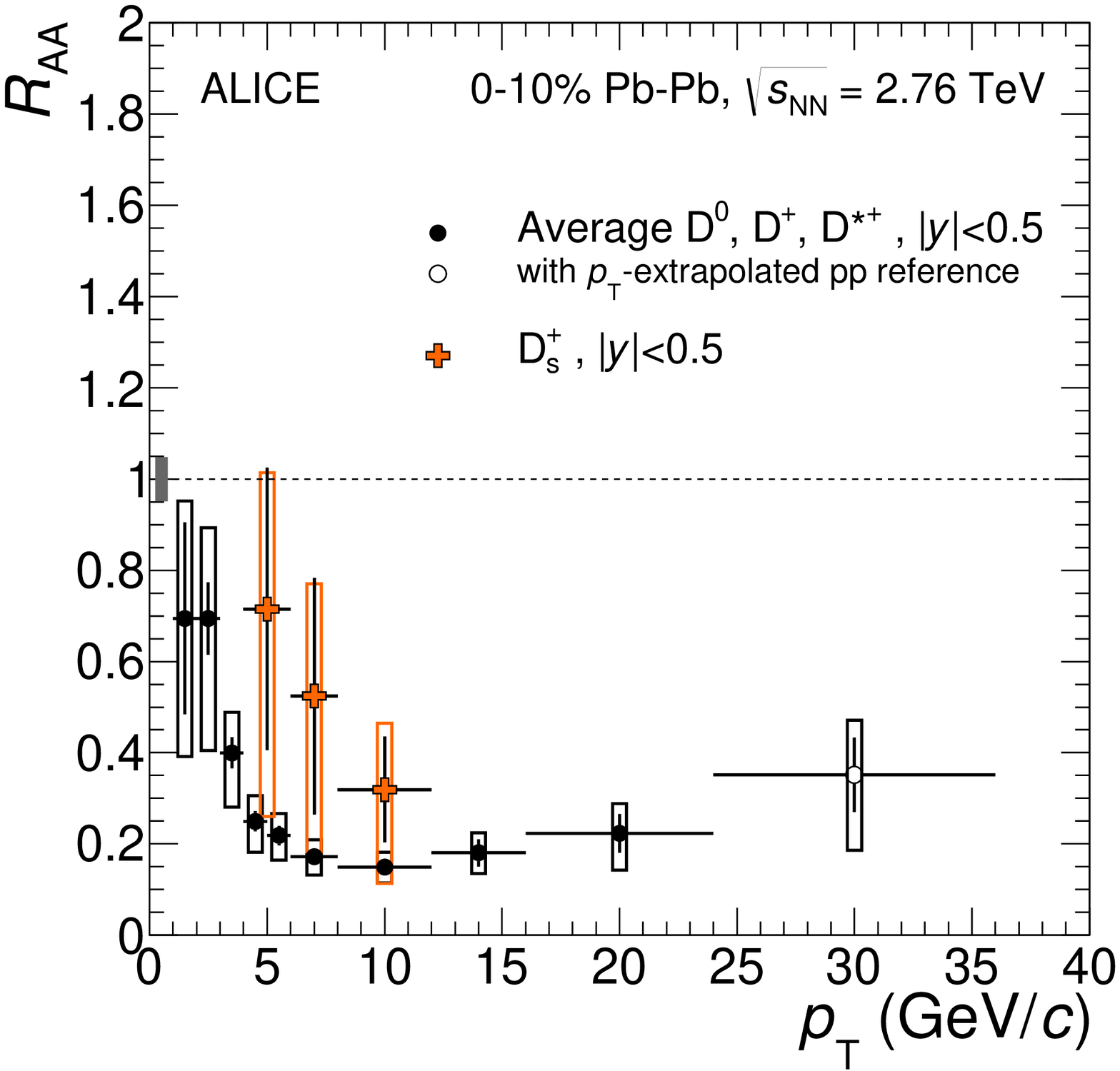}
\caption{Left: $\RAA$ of prompt $\Ds$ mesons in the 0--10\% and 20--50\% 
centrality classes as a function of $\pt$. For the 20--50\% case, the symbols 
are displaced horizontally for visibility.
Right: $\RAA$ of prompt $\Ds$ mesons compared to non-strange D mesons 
(average of $\Dzero$, $\Dplus$ and $\Dstar$~\cite{Adam:2015sza}) in the 
0--10\% centrality class.
Statistical (bars), systematic (empty boxes), and normalisation (full box) 
uncertainties are shown.}
\label{fig:RAApt}
\end{center}
\end{figure}

The nuclear modification factor $\RAA$ of prompt $\Ds$ mesons was computed from
the $\mathrm{d} N/\mathrm{d}\pt$ distributions.
The results are shown as a function of $\pt$ in the left-hand panel of 
Fig.~\ref{fig:RAApt} for the two centrality classes.
The vertical bars represent the statistical uncertainties, the empty
boxes are the total $\pt$-dependent systematic uncertainties described in
Section~\ref{sec:systematics},  except for the normalisation uncertainty, which 
is displayed as a filled box at $\RAA=1$.
A suppression by a factor of about three of the $\Ds$-meson yield 
in Pb--Pb collisions relative to the binary-scaled pp cross section is observed
in the highest $\pt$ interval ($8<\pt<12~\gev/c$) for the 10\% most central
collisions.
A smaller suppression (by a factor of about two) is measured in the 
20--50\% centrality class in $8<\pt<12~\gev/c$, even though
with the current uncertainties no conclusions can be drawn on the 
centrality dependence of the $\Ds$-meson nuclear modification factor at high $\pt$.
Since no significant modification of the $\Ds$-meson production relative 
to binary-scaled pp collisions is observed in p--Pb 
reactions in the $\pt$ range considered here~\cite{ALICEDRpPb},
the substantial suppression of the $\Ds$-meson yield at high $\pt$ 
in Pb--Pb collisions cannot be explained in terms of initial state effects, 
but it is predominantly due to strong final-state effects induced by the
hot and dense partonic medium created in the collisions of heavy nuclei.
At lower $\pt$ the central values of the measurement show a larger
$\RAA$, however the large statistical and systematic
uncertainties do not allow to draw a conclusion on the $\pt$ dependence
of the $\Ds$ nuclear modification factor.

The $\RAA$ of prompt $\Ds$ mesons in the
10\% most central collisions is compared in the right-hand panel of 
Fig.~\ref{fig:RAApt} to the average nuclear modification factor of
 $\Dzero$, $\Dplus$ and $\Dstar$ mesons measured in the same centrality 
class~\cite{Adam:2015sza}. This comparison is meant to address the expected effect of hadronisation via quark 
recombination in the partonic medium on the relative abundances of strange
and non-strange D-meson species.
In the three $\pt$ intervals, the values of
the $\Ds$-meson $\RAA$ are higher than those of non-strange D mesons, although 
compatible within uncertainties.
Even considering that a part of the systematic uncertainty is correlated
between strange and non-strange D mesons, the current uncertainties do not 
allow a conclusive statement on the expected enhancement of the $\Ds$-meson 
yield relative to that of non-strange D mesons in heavy-ion collisions.

\begin{figure}[!t]
\begin{center}
\includegraphics[width=0.48\textwidth]{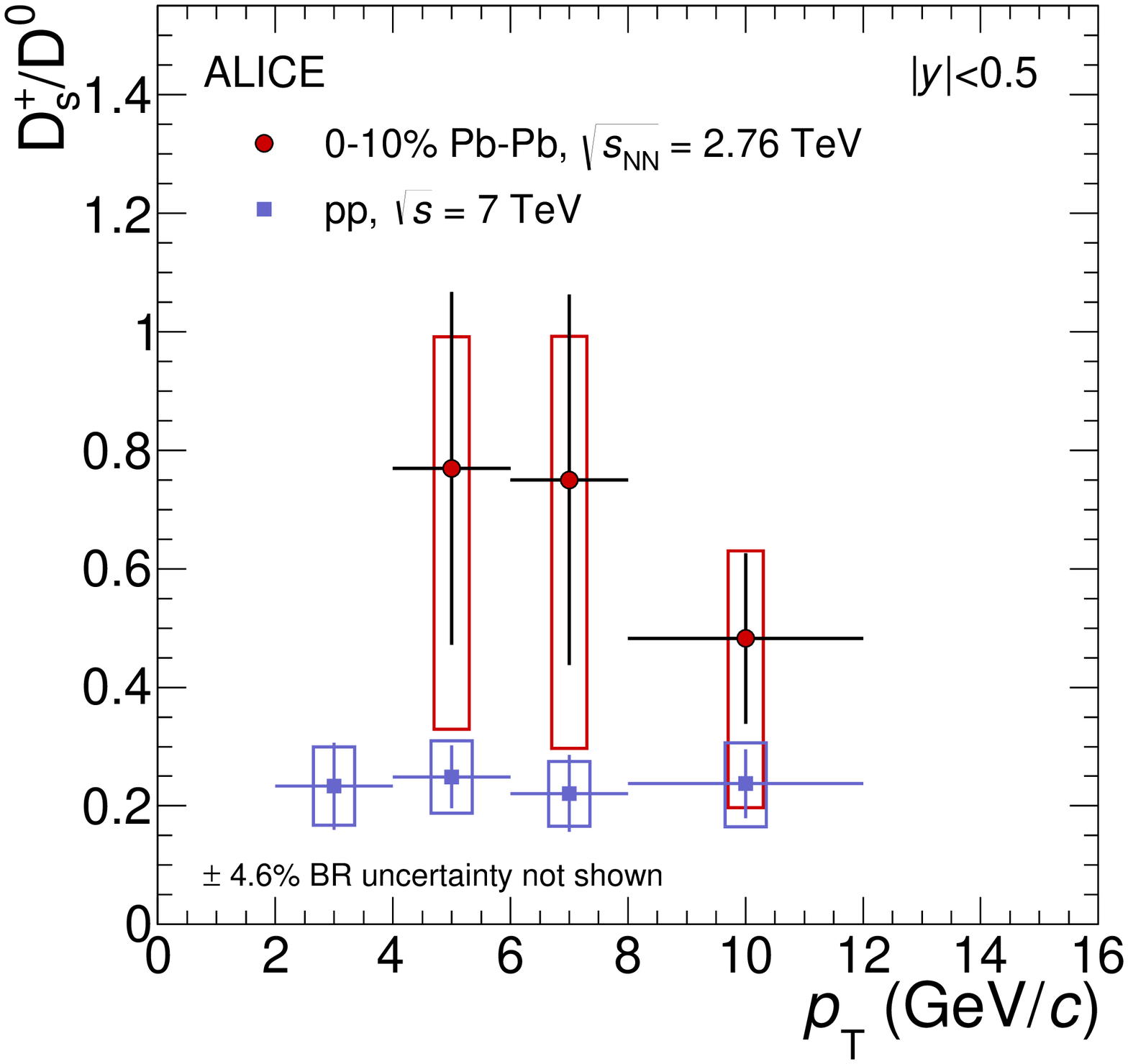}
\includegraphics[width=0.48\textwidth]{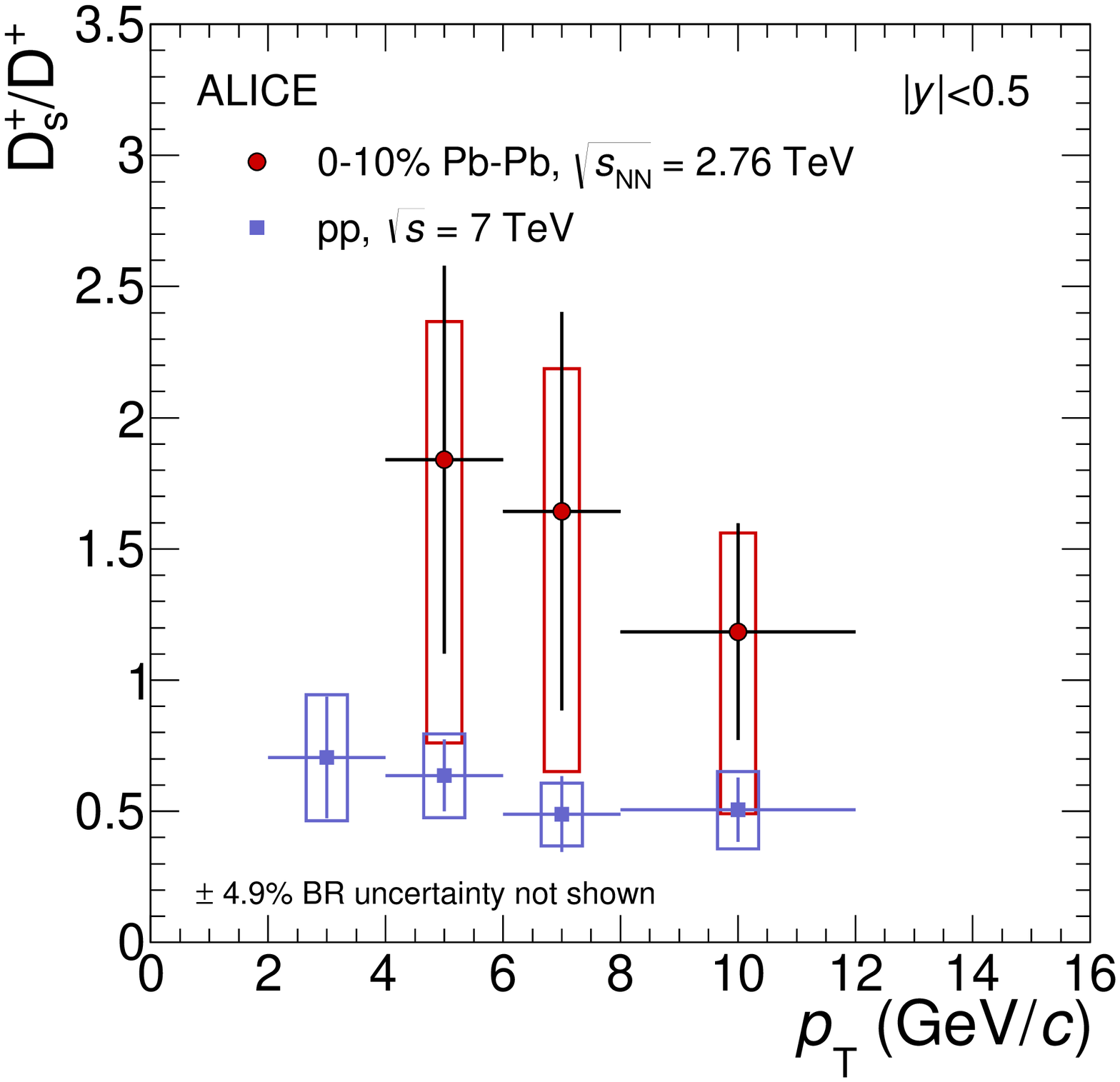}
\caption{Ratios of prompt D-meson yields ($\Ds$/$\Dzero$ and $\Ds$/$\Dplus$) 
as a function of $\pt$ in the 10\% most central 
Pb--Pb collisions at $\sqrtsNN=2.76~\tev$ compared to the results
in pp collisions at $\sqrt{s}=7~\TeV$. Statistical (bars) and systematic (boxes) uncertainties are shown. }
\label{fig:ratiosPbPb}
\end{center}
\end{figure}

An alternative approach to study the predicted modification of the charm-quark 
hadronisation in the presence of a QGP is to compare the ratios 
between the measured yields of $\Ds$ and $\Dzero$($\Dplus$) mesons in 
Pb--Pb and pp collisions.
This comparison is shown in Fig.~\ref{fig:ratiosPbPb} for the 10\% most 
central Pb--Pb collisions. In the left-hand panel the $\Ds/\Dzero$ ratio is displayed, while 
the right-hand panel shows the ratio $\Ds/\Dplus$.
The ratios $\Ds/\Dzero$ and $\Ds/\Dplus$ in pp collisions 
are taken from the measurements at 
$\sqrt{s}=7~\TeV$~\cite{ALICEDspp7}\footnote{
The values from Ref.~\cite{ALICEDspp7} were re-computed
with the most recent value for the branching ratio of the 
$\DstophipitoKKpi$ decay chain, which is 2.24\%~\cite{PDG}, while
it was 2.28\% at the time of the pp publication.
}. 
No strong dependence on the collision energy is expected 
(see~\cite{ALICEDspp7} and references therein).
In the evaluation of the systematic uncertainties on the D-meson yield ratios, 
the sources of correlated and uncorrelated systematic effects were treated 
separately. 
In particular, the contributions of the yield extraction, topological selection efficiency and 
PID efficiency were considered as uncorrelated and summed in quadrature.
The uncertainty on the tracking efficiency cancels completely in the
ratios between production cross sections of meson species reconstructed from 
three-body decay channels ($\Dplus$ and $\Ds$), while a 5\% 
systematic uncertainty (4\% in the pp case) was considered in the ratio 
to the $\Dzero$ yields, which are reconstructed from a two-particle final 
state.
To propagate the uncertainty due to the B feed-down subtraction,
the contribution of the FONLL cross section was treated as completely
correlated among the D-meson species. It was estimated from the spread of 
the D-meson yield ratios obtained by varying the factorisation and 
renormalisation scales and the heavy-quark mass in FONLL 
coherently for the three meson species.
The contribution due to the hypothesis on 
$R_{{\rm AA}}^{\textnormal{feed-down}}/R_{{\rm AA}}^{\rm prompt}$ was considered as 
uncorrelated between $\Ds$ and non-strange D mesons and summed in quadrature.
The difference between the $\Ds/\Dzero$ ratios in pp and in central Pb--Pb 
collisions is of about $1\,\sigma$ of the combined statistical and systematic 
uncertainties in both the two lowest $\pt$ intervals, 
$4<\pt<6~\gev/c$ and $6<\pt<8~\gev/c$.
An enhancement of D$_{\rm s}$/D ratios in heavy-ion collisions
is predicted if recombination contributes to charm 
quark hadronisation in the QGP.
However, considering the current level of experimental uncertainties, 
no conclusion on charm-quark hadronisation can be drawn from this
first measurement of $\Ds$-meson production in Pb--Pb collisions.

In the framework of the Statistical Hadronisation 
Model~\cite{Andronic2003,SHM1,SHM2}, the $\pt$-integrated ratios of D-meson 
abundances for a chemical freeze-out temperature $T=156~\MeV$ (as extracted 
from fits to the measured abundances of light-flavour 
hadrons~\cite{Stachel:2013zma}) and vanishing baryo-chemical potential, are 
expected to be $\Ds/\Dzero=0.338$ and $\Ds/\Dplus=0.830$, which
are higher by a factor of about two with respect to the values calculated for 
pp collisions at LHC energies~\cite{ALICEDspp7}.

\begin{figure}[!t]
\begin{center}
\includegraphics[width=0.48\textwidth]{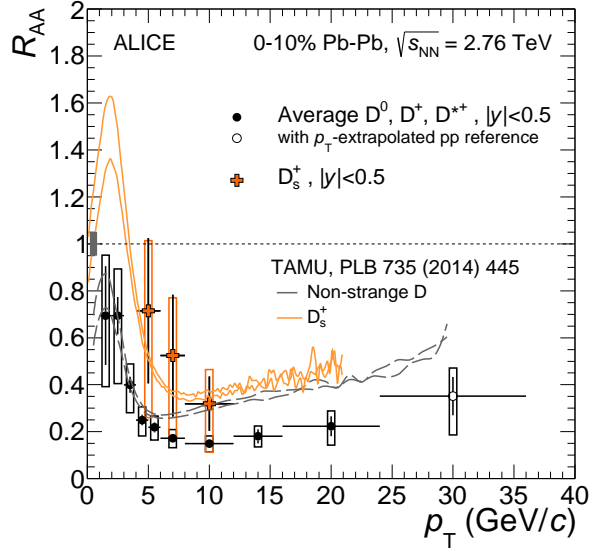}
\caption{$\RAA$ of prompt $\Ds$ and non-strange D mesons 
(average of $\Dzero$, $\Dplus$ and $\Dstar$) in the 0--10\% centrality class
compared to predictions of the TAMU model~\cite{TAMULHC}.
The bands shown for the TAMU predictions encompass the charm-shadowing 
uncertainty.}
\label{fig:RAAvsmodels}
\end{center}
\end{figure}

In Fig.~\ref{fig:RAAvsmodels}, the measured $\RAA$ of non-strange D mesons
and of $\Ds$ are compared to the prediction of the TAMU 
model~\cite{TAMU,TAMULHC}.
Among the several models available for open charm production in heavy-ion 
collisions, TAMU is the only one providing a quantitative prediction for the 
$\Ds$-meson nuclear modification factor.
This is a heavy-quark transport model based on heavy-quark diffusion,
implemented via simulations based on the relativistic Langevin equation, 
in a hydrodynamically expanding medium.
The interactions of the charm quarks with the medium are modeled including
only elastic processes, which are assumed to govern 
the heavy-quark scattering amplitudes at low and intermediate momenta.
The heavy-quark transport coefficients are calculated within a 
non-perturbative $T$-matrix approach, where the interactions proceed via 
resonance formation that transfers momentum from the 
heavy quarks to the medium constituents. 
The hadronisation of charm quarks is performed via recombination with 
thermalized up, down and strange quarks.
The remaining charm quarks are converted to hadrons using the vacuum 
fragmentation functions from~\cite{Braaten} and fragmentation 
fractions $f({\rm c\rightarrow D})$ from PYTHIA.
This model predicts an enhancement of the $\Ds$ over the non-strange D-meson 
$\RAA$
at low $\pt$ as a consequence of the recombination of charm quarks with 
thermally equilibrated strange quarks in the QGP.
At higher $\pt$, where the dominant hadronisation mechanism is fragmentation,
similar $\RAA$ values are predicted for the different D-meson species.
The model describes the measured $\Ds$-meson nuclear modification factor within
uncertainties and at low $\pt$ provides also a reasonable description of 
non-strange D-meson $\RAA$. 
The measured suppression of non-strange D mesons is underestimated at 
higher $\pt$, 
where the contribution of inelastic processes (gluon radiation),
which are missing in this transport calculation, is expected to play a major 
role.

\section{Summary}
\label{sec:summary}
The production of $\Ds$ mesons was measured for the first time in
heavy-ion collisions. The measurement was carried out on a sample of
Pb--Pb collisions at $\sqrtsNN=2.76~\tev$ in two centrality classes, namely 0--10\%
and 20--50\%.

The results for the 10\% most central collisions 
indicate a substantial suppression ($\RAA\approx0.3$) of the production of 
$\Ds$ mesons at high $\pt$ ($8<\pt<12~\gev/c$) with respect to the expectation 
based on the pp cross section scaled by the average nuclear overlap function.
The observed suppression is compatible with that of non-strange D mesons and
can be described by models including strong coupling of the charm quarks
with the deconfined medium formed in the collision.

At lower momenta ($4<\pt<8~\gev/c$), the values of the $\Ds$-meson nuclear
modification factor are larger than those of non-strange D mesons,
although compatible within uncertainties.
This result provides a possible hint for an enhancement of D$_{\rm s}$/D 
ratio, which is expected if the recombination process
significantly contributes to the charm quark hadronisation in the QGP.

The precision of the measurements will be improved using the
larger data samples of Pb--Pb collisions that will be collected
during the ongoing LHC Run-2.
The larger sample size will allow us to observe the $\Ds$ signal with less 
stringent selections, thus reducing the systematic uncertainty on the 
efficiency correction.
In addition, the higher Pb--Pb collision centre-of-mass energy will reduce
the impact of the $\sqrt{s}$-scaling of the pp reference.
This will open the possibility to exploit the measurement of $\Ds$-meson
production in heavy-ion collisions to assess
the recombination effects in the charm-quark hadronisation and to provide
further constraints to models describing the coupling of heavy quarks with 
the medium.

\newenvironment{acknowledgement}{\relax}{\relax}
\begin{acknowledgement}
\section*{Acknowledgements}

The ALICE Collaboration would like to thank all its engineers and technicians for their invaluable contributions to the construction of the experiment and the CERN accelerator teams for the outstanding performance of the LHC complex.
The ALICE Collaboration gratefully acknowledges the resources and support provided by all Grid centres and the Worldwide LHC Computing Grid (WLCG) collaboration.
The ALICE Collaboration would like to thank M. He, R. Fries and R. Rapp
for making available their model calculations.
The ALICE Collaboration acknowledges the following funding agencies for their support in building and
running the ALICE detector:
State Committee of Science,  World Federation of Scientists (WFS)
and Swiss Fonds Kidagan, Armenia;
Conselho Nacional de Desenvolvimento Cient\'{\i}fico e Tecnol\'{o}gico (CNPq), Financiadora de Estudos e Projetos (FINEP),
Funda\c{c}\~{a}o de Amparo \`{a} Pesquisa do Estado de S\~{a}o Paulo (FAPESP);
National Natural Science Foundation of China (NSFC), the Chinese Ministry of Education (CMOE)
and the Ministry of Science and Technology of China (MSTC);
Ministry of Education and Youth of the Czech Republic;
Danish Natural Science Research Council, the Carlsberg Foundation and the Danish National Research Foundation;
The European Research Council under the European Community's Seventh Framework Programme;
Helsinki Institute of Physics and the Academy of Finland;
French CNRS-IN2P3, the `Region Pays de Loire', `Region Alsace', `Region Auvergne' and CEA, France;
German Bundesministerium fur Bildung, Wissenschaft, Forschung und Technologie (BMBF) and the Helmholtz Association;
General Secretariat for Research and Technology, Ministry of Development, Greece;
Hungarian Orszagos Tudomanyos Kutatasi Alappgrammok (OTKA) and National Office for Research and Technology (NKTH);
Department of Atomic Energy and Department of Science and Technology of the Government of India;
Istituto Nazionale di Fisica Nucleare (INFN) and Centro Fermi -
Museo Storico della Fisica e Centro Studi e Ricerche ``Enrico Fermi'', Italy;
MEXT Grant-in-Aid for Specially Promoted Research, Ja\-pan;
Joint Institute for Nuclear Research, Dubna;
National Research Foundation of Korea (NRF);
Consejo Nacional de Cienca y Tecnologia (CONACYT), Direccion General de Asuntos del Personal Academico(DGAPA), M\'{e}xico, Amerique Latine Formation academique - European Commission~(ALFA-EC) and the EPLANET Program~(European Particle Physics Latin American Network);
Stichting voor Fundamenteel Onderzoek der Materie (FOM) and the Nederlandse Organisatie voor Wetenschappelijk Onderzoek (NWO), Netherlands;
Research Council of Norway (NFR);
National Science Centre, Poland;
Ministry of National Education/Institute for Atomic Physics and National Council of Scientific Research in Higher Education~(CNCSI-UEFISCDI), Romania;
Ministry of Education and Science of Russian Federation, Russian
Academy of Sciences, Russian Federal Agency of Atomic Energy,
Russian Federal Agency for Science and Innovations and The Russian
Foundation for Basic Research;
Ministry of Education of Slovakia;
Department of Science and Technology, South Africa;
Centro de Investigaciones Energeticas, Medioambientales y Tecnologicas (CIEMAT), E-Infrastructure shared between Europe and Latin America (EELA), Ministerio de Econom\'{i}a y Competitividad (MINECO) of Spain, Xunta de Galicia (Conseller\'{\i}a de Educaci\'{o}n),
Centro de Aplicaciones Tecnológicas y Desarrollo Nuclear (CEA\-DEN), Cubaenerg\'{\i}a, Cuba, and IAEA (International Atomic Energy Agency);
Swedish Research Council (VR) and Knut $\&$ Alice Wallenberg
Foundation (KAW);
Ukraine Ministry of Education and Science;
United Kingdom Science and Technology Facilities Council (STFC);
The United States Department of Energy, the United States National
Science Foundation, the State of Texas, and the State of Ohio;
Ministry of Science, Education and Sports of Croatia and  Unity through Knowledge Fund, Croatia;
Council of Scientific and Industrial Research (CSIR), New Delhi, India;
Pontificia Universidad Cat\'{o}lica del Per\'{u}.
\end{acknowledgement}

\bibliographystyle{utphys}
\bibliography{biblio}
%

%

\newpage
\appendix
\section{The ALICE Collaboration}
\label{app:collab}



\begingroup
\small
\begin{flushleft}
J.~Adam\Irefn{org40}\And
D.~Adamov\'{a}\Irefn{org83}\And
M.M.~Aggarwal\Irefn{org87}\And
G.~Aglieri Rinella\Irefn{org36}\And
M.~Agnello\Irefn{org110}\And
N.~Agrawal\Irefn{org48}\And
Z.~Ahammed\Irefn{org132}\And
S.U.~Ahn\Irefn{org68}\And
S.~Aiola\Irefn{org136}\And
A.~Akindinov\Irefn{org58}
% ...
\renewcommand\labelenumi{\textsuperscript{\theenumi}~}

\section*{Affiliation notes}
\renewcommand\theenumi{\roman{enumi}}
\begin{Authlist}
\item \Adef{0}Deceased
\item \Adef{idp1748160}{Also at: Georgia State University, Atlanta, Georgia, United States}
\item \Adef{idp3790688}{Also at: M.V. Lomonosov Moscow State University, D.V. Skobeltsyn Institute of Nuclear, Physics, Moscow, Russia}
\end{Authlist}

\section*{Collaboration Institutes}
\renewcommand\theenumi{\arabic{enumi}~}
\begin{Authlist}

\item \Idef{org1}A.I. Alikhanyan National Science Laboratory (Yerevan Physics Institute) Foundation, Yerevan, Armenia
\item \Idef{org2}Benem\'{e}rita Universidad Aut\'{o}noma de Puebla, Puebla, Mexico
\item \Idef{org3}Bogolyubov Institute for Theoretical Physics, Kiev, Ukraine
\item \Idef{org4}Bose Institute, Department of Physics and Centre for Astroparticle Physics and Space Science (CAPSS), Kolkata, India
\item \Idef{org5}Budker Institute for Nuclear Physics, Novosibirsk, Russia
\item \Idef{org6}California Polytechnic State University, San Luis Obispo, California, United States
\item \Idef{org7}Central China Normal University, Wuhan, China
\item \Idef{org8}Centre de Calcul de l'IN2P3, Villeurbanne, France
\item \Idef{org9}Centro de Aplicaciones Tecnol\'{o}gicas y Desarrollo Nuclear (CEADEN), Havana, Cuba
\item \Idef{org10}Centro de Investigaciones Energ\'{e}ticas Medioambientales y Tecnol\'{o}gicas (CIEMAT), Madrid, Spain
\item \Idef{org11}Centro de Investigaci\'{o}n y de Estudios Avanzados (CINVESTAV), Mexico City and M\'{e}rida, Mexico
\item \Idef{org12}Centro Fermi - Museo Storico della Fisica e Centro Studi e Ricerche ``Enrico Fermi'', Rome, Italy
\item \Idef{org13}Chicago State University, Chicago, Illinois, USA
\item \Idef{org14}China Institute of Atomic Energy, Beijing, China
\item \Idef{org15}Commissariat \`{a} l'Energie Atomique, IRFU, Saclay, France
\item \Idef{org16}COMSATS Institute of Information Technology (CIIT), Islamabad, Pakistan
\item \Idef{org17}Departamento de F\'{\i}sica de Part\'{\i}culas and IGFAE, Universidad de Santiago de Compostela, Santiago de Compostela, Spain
\item \Idef{org18}Department of Physics and Technology, University of Bergen, Bergen, Norway
\item \Idef{org19}Department of Physics, Aligarh Muslim University, Aligarh, India
\item \Idef{org20}Department of Physics, Ohio State University, Columbus, Ohio, United States
\item \Idef{org21}Department of Physics, Sejong University, Seoul, South Korea
\item \Idef{org22}Department of Physics, University of Oslo, Oslo, Norway
\item \Idef{org23}Dipartimento di Elettrotecnica ed Elettronica del Politecnico, Bari, Italy
\item \Idef{org24}Dipartimento di Fisica dell'Universit\`{a} 'La Sapienza' and Sezione INFN Rome, Italy
\item \Idef{org25}Dipartimento di Fisica dell'Universit\`{a} and Sezione INFN, Cagliari, Italy
\item \Idef{org26}Dipartimento di Fisica dell'Universit\`{a} and Sezione INFN, Trieste, Italy
\item \Idef{org27}Dipartimento di Fisica dell'Universit\`{a} and Sezione INFN, Turin, Italy
\item \Idef{org28}Dipartimento di Fisica e Astronomia dell'Universit\`{a} and Sezione INFN, Bologna, Italy
\item \Idef{org29}Dipartimento di Fisica e Astronomia dell'Universit\`{a} and Sezione INFN, Catania, Italy
\item \Idef{org30}Dipartimento di Fisica e Astronomia dell'Universit\`{a} and Sezione INFN, Padova, Italy
\item \Idef{org31}Dipartimento di Fisica `E.R.~Caianiello' dell'Universit\`{a} and Gruppo Collegato INFN, Salerno, Italy
\item \Idef{org32}Dipartimento di Scienze e Innovazione Tecnologica dell'Universit\`{a} del  Piemonte Orientale and Gruppo Collegato INFN, Alessandria, Italy
\item \Idef{org33}Dipartimento Interateneo di Fisica `M.~Merlin' and Sezione INFN, Bari, Italy
\item \Idef{org34}Division of Experimental High Energy Physics, University of Lund, Lund, Sweden
\item \Idef{org35}Eberhard Karls Universit\"{a}t T\"{u}bingen, T\"{u}bingen, Germany
\item \Idef{org36}European Organization for Nuclear Research (CERN), Geneva, Switzerland
\item \Idef{org37}Excellence Cluster Universe, Technische Universit\"{a}t M\"{u}nchen, Munich, Germany
\item \Idef{org38}Faculty of Engineering, Bergen University College, Bergen, Norway
\item \Idef{org39}Faculty of Mathematics, Physics and Informatics, Comenius University, Bratislava, Slovakia
\item \Idef{org40}Faculty of Nuclear Sciences and Physical Engineering, Czech Technical University in Prague, Prague, Czech Republic
\item \Idef{org41}Faculty of Science, P.J.~\v{S}af\'{a}rik University, Ko\v{s}ice, Slovakia
\item \Idef{org42}Faculty of Technology, Buskerud and Vestfold University College, Vestfold, Norway
\item \Idef{org43}Frankfurt Institute for Advanced Studies, Johann Wolfgang Goethe-Universit\"{a}t Frankfurt, Frankfurt, Germany
\item \Idef{org44}Gangneung-Wonju National University, Gangneung, South Korea
\item \Idef{org45}Gauhati University, Department of Physics, Guwahati, India
\item \Idef{org46}Helsinki Institute of Physics (HIP), Helsinki, Finland
\item \Idef{org47}Hiroshima University, Hiroshima, Japan
\item \Idef{org48}Indian Institute of Technology Bombay (IIT), Mumbai, India
\item \Idef{org49}Indian Institute of Technology Indore, Indore (IITI), India
\item \Idef{org50}Inha University, Incheon, South Korea
\item \Idef{org51}Institut de Physique Nucl\'eaire d'Orsay (IPNO), Universit\'e Paris-Sud, CNRS-IN2P3, Orsay, France
\item \Idef{org52}Institut f\"{u}r Informatik, Johann Wolfgang Goethe-Universit\"{a}t Frankfurt, Frankfurt, Germany
\item \Idef{org53}Institut f\"{u}r Kernphysik, Johann Wolfgang Goethe-Universit\"{a}t Frankfurt, Frankfurt, Germany
\item \Idef{org54}Institut f\"{u}r Kernphysik, Westf\"{a}lische Wilhelms-Universit\"{a}t M\"{u}nster, M\"{u}nster, Germany
\item \Idef{org55}Institut Pluridisciplinaire Hubert Curien (IPHC), Universit\'{e} de Strasbourg, CNRS-IN2P3, Strasbourg, France
\item \Idef{org56}Institute for Nuclear Research, Academy of Sciences, Moscow, Russia
\item \Idef{org57}Institute for Subatomic Physics of Utrecht University, Utrecht, Netherlands
\item \Idef{org58}Institute for Theoretical and Experimental Physics, Moscow, Russia
\item \Idef{org59}Institute of Experimental Physics, Slovak Academy of Sciences, Ko\v{s}ice, Slovakia
\item \Idef{org60}Institute of Physics, Academy of Sciences of the Czech Republic, Prague, Czech Republic
\item \Idef{org61}Institute of Physics, Bhubaneswar, India
\item \Idef{org62}Institute of Space Science (ISS), Bucharest, Romania
\item \Idef{org63}Instituto de Ciencias Nucleares, Universidad Nacional Aut\'{o}noma de M\'{e}xico, Mexico City, Mexico
\item \Idef{org64}Instituto de F\'{\i}sica, Universidad Nacional Aut\'{o}noma de M\'{e}xico, Mexico City, Mexico
\item \Idef{org65}iThemba LABS, National Research Foundation, Somerset West, South Africa
\item \Idef{org66}Joint Institute for Nuclear Research (JINR), Dubna, Russia
\item \Idef{org67}Konkuk University, Seoul, South Korea
\item \Idef{org68}Korea Institute of Science and Technology Information, Daejeon, South Korea
\item \Idef{org69}KTO Karatay University, Konya, Turkey
\item \Idef{org70}Laboratoire de Physique Corpusculaire (LPC), Clermont Universit\'{e}, Universit\'{e} Blaise Pascal, CNRS--IN2P3, Clermont-Ferrand, France
\item \Idef{org71}Laboratoire de Physique Subatomique et de Cosmologie, Universit\'{e} Grenoble-Alpes, CNRS-IN2P3, Grenoble, France
\item \Idef{org72}Laboratori Nazionali di Frascati, INFN, Frascati, Italy
\item \Idef{org73}Laboratori Nazionali di Legnaro, INFN, Legnaro, Italy
\item \Idef{org74}Lawrence Berkeley National Laboratory, Berkeley, California, United States
\item \Idef{org75}Moscow Engineering Physics Institute, Moscow, Russia
\item \Idef{org76}Nagasaki Institute of Applied Science, Nagasaki, Japan
\item \Idef{org77}National Centre for Nuclear Studies, Warsaw, Poland
\item \Idef{org78}National Institute for Physics and Nuclear Engineering, Bucharest, Romania
\item \Idef{org79}National Institute of Science Education and Research, Bhubaneswar, India
\item \Idef{org80}Niels Bohr Institute, University of Copenhagen, Copenhagen, Denmark
\item \Idef{org81}Nikhef, Nationaal instituut voor subatomaire fysica, Amsterdam, Netherlands
\item \Idef{org82}Nuclear Physics Group, STFC Daresbury Laboratory, Daresbury, United Kingdom
\item \Idef{org83}Nuclear Physics Institute, Academy of Sciences of the Czech Republic, \v{R}e\v{z} u Prahy, Czech Republic
\item \Idef{org84}Oak Ridge National Laboratory, Oak Ridge, Tennessee, United States
\item \Idef{org85}Petersburg Nuclear Physics Institute, Gatchina, Russia
\item \Idef{org86}Physics Department, Creighton University, Omaha, Nebraska, United States
\item \Idef{org87}Physics Department, Panjab University, Chandigarh, India
\item \Idef{org88}Physics Department, University of Athens, Athens, Greece
\item \Idef{org89}Physics Department, University of Cape Town, Cape Town, South Africa
\item \Idef{org90}Physics Department, University of Jammu, Jammu, India
\item \Idef{org91}Physics Department, University of Rajasthan, Jaipur, India
\item \Idef{org92}Physik Department, Technische Universit\"{a}t M\"{u}nchen, Munich, Germany
\item \Idef{org93}Physikalisches Institut, Ruprecht-Karls-Universit\"{a}t Heidelberg, Heidelberg, Germany
\item \Idef{org94}Purdue University, West Lafayette, Indiana, United States
\item \Idef{org95}Pusan National University, Pusan, South Korea
\item \Idef{org96}Research Division and ExtreMe Matter Institute EMMI, GSI Helmholtzzentrum f\"ur Schwerionenforschung, Darmstadt, Germany
\item \Idef{org97}Rudjer Bo\v{s}kovi\'{c} Institute, Zagreb, Croatia
\item \Idef{org98}Russian Federal Nuclear Center (VNIIEF), Sarov, Russia
\item \Idef{org99}Russian Research Centre Kurchatov Institute, Moscow, Russia
\item \Idef{org100}Saha Institute of Nuclear Physics, Kolkata, India
\item \Idef{org101}School of Physics and Astronomy, University of Birmingham, Birmingham, United Kingdom
\item \Idef{org102}Secci\'{o}n F\'{\i}sica, Departamento de Ciencias, Pontificia Universidad Cat\'{o}lica del Per\'{u}, Lima, Peru
\item \Idef{org103}Sezione INFN, Bari, Italy
\item \Idef{org104}Sezione INFN, Bologna, Italy
\item \Idef{org105}Sezione INFN, Cagliari, Italy
\item \Idef{org106}Sezione INFN, Catania, Italy
\item \Idef{org107}Sezione INFN, Padova, Italy
\item \Idef{org108}Sezione INFN, Rome, Italy
\item \Idef{org109}Sezione INFN, Trieste, Italy
\item \Idef{org110}Sezione INFN, Turin, Italy
\item \Idef{org111}SSC IHEP of NRC Kurchatov institute, Protvino, Russia
\item \Idef{org112}Stefan Meyer Institut f\"{u}r Subatomare Physik (SMI), Vienna, Austria
\item \Idef{org113}SUBATECH, Ecole des Mines de Nantes, Universit\'{e} de Nantes, CNRS-IN2P3, Nantes, France
\item \Idef{org114}Suranaree University of Technology, Nakhon Ratchasima, Thailand
\item \Idef{org115}Technical University of Ko\v{s}ice, Ko\v{s}ice, Slovakia
\item \Idef{org116}Technical University of Split FESB, Split, Croatia
\item \Idef{org117}The Henryk Niewodniczanski Institute of Nuclear Physics, Polish Academy of Sciences, Cracow, Poland
\item \Idef{org118}The University of Texas at Austin, Physics Department, Austin, Texas, USA
\item \Idef{org119}Universidad Aut\'{o}noma de Sinaloa, Culiac\'{a}n, Mexico
\item \Idef{org120}Universidade de S\~{a}o Paulo (USP), S\~{a}o Paulo, Brazil
\item \Idef{org121}Universidade Estadual de Campinas (UNICAMP), Campinas, Brazil
\item \Idef{org122}University of Houston, Houston, Texas, United States
\item \Idef{org123}University of Jyv\"{a}skyl\"{a}, Jyv\"{a}skyl\"{a}, Finland
\item \Idef{org124}University of Liverpool, Liverpool, United Kingdom
\item \Idef{org125}University of Tennessee, Knoxville, Tennessee, United States
\item \Idef{org126}University of the Witwatersrand, Johannesburg, South Africa
\item \Idef{org127}University of Tokyo, Tokyo, Japan
\item \Idef{org128}University of Tsukuba, Tsukuba, Japan
\item \Idef{org129}University of Zagreb, Zagreb, Croatia
\item \Idef{org130}Universit\'{e} de Lyon, Universit\'{e} Lyon 1, CNRS/IN2P3, IPN-Lyon, Villeurbanne, France
\item \Idef{org131}V.~Fock Institute for Physics, St. Petersburg State University, St. Petersburg, Russia
\item \Idef{org132}Variable Energy Cyclotron Centre, Kolkata, India
\item \Idef{org133}Warsaw University of Technology, Warsaw, Poland
\item \Idef{org134}Wayne State University, Detroit, Michigan, United States
\item \Idef{org135}Wigner Research Centre for Physics, Hungarian Academy of Sciences, Budapest, Hungary
\item \Idef{org136}Yale University, New Haven, Connecticut, United States
\item \Idef{org137}Yonsei University, Seoul, South Korea
\item \Idef{org138}Zentrum f\"{u}r Technologietransfer und Telekommunikation (ZTT), Fachhochschule Worms, Worms, Germany
\end{Authlist}
\endgroup

\end{document}